\documentclass[prf,twocolumn,amsmath,amssymb,floatfix,showpacs]{revtex4-1}

\usepackage{tikz}
\usepackage{graphicx}
\usepackage{dcolumn}
\usepackage{bm}
\usepackage{mciteplus}
\usepackage{braket}

\usepackage{newfloat,algcompatible}
\usepackage[size=small]{caption}
\usepackage{etoolbox}

\usepackage[normalem]{ulem}

\DeclareFloatingEnvironment[
    fileext=loa,
    listname=List of Algorithms,
    name=ALGORITHM,
    placement=tbhp,
]{algorithm}
\DeclareCaptionFormat{algorithms}{\vskip-15pt\hrulefill\par#1#2#3\vskip-6pt\hrulefill}
\usepackage{shortCut}
\graphicspath{{PRFeps/}}
\begin{document}

\title{Large-scale instabilities of helical flows}
\author{Alexandre \textsc{Cameron}}
\email[]{alexandre.cameron@ens.fr}
\author{Alexandros \textsc{Alexakis}} 
\email[]{alexakis@lps.ens.fr}
\author{Marc-{\'E}tienne~\textsc{Brachet}}
\email[]{brachet@physique.ens.fr}
\affiliation{Laboratoire de Physique Statistique, 
  \'Ecole Normale Sup\'erieure, PSL Research University; 
  Universit\'e Paris Diderot Sorbonne Paris-Cit\'e; 
  Sorbonne Universit\'es UPMC Univ Paris 06; CNRS; 
  24 rue Lhomond, 75005 Paris, France}
\date{\today}
%%%%%%%%%%%%%%%%%%%%%%%%%%%%%%%%%%%%
\begin{abstract}
Large-scale hydrodynamic instabilities of periodic helical flows are investigated using 
$3$D Floquet numerical computations. A minimal three-modes analytical model that reproduce 
and explains some of the full Floquet results is derived. The growth-rate $\sigma$ of the 
most unstable modes (at small scale, low Reynolds number $Re$ and small wavenumber $q$) 
is found to scale differently in the presence or absence of anisotropic kinetic alpha (\AKA{}) effect.
When an \AKA{} effect is present the scaling $\sigma \propto q\; Re\,$ predicted by the \AKA{} 
effect theory 
[U. Frisch, Z. S. She, and P. L. Sulem, Physica D: Nonlinear Phenomena 28, 382 (1987)]
is recovered for $Re\ll 1$ as expected (with most of the energy of the unstable mode 
concentrated in the large scales). 
However, as $Re$ increases, the growth-rate is found to saturate and most of the 
energy is found at small scales. In the absence of \AKA{} effect, it is found that flows 
can still have large-scale instabilities, but with a negative eddy-viscosity scaling 
$\sigma \propto \nu(b Re^2-1) q^2$. The instability appears only above a critical value 
of the Reynolds number $\rec$. For values of $Re$ above a second critical value $\recS$ beyond
which small-scale instabilities are present, the growth-rate becomes independent of $q$ and 
the energy of the perturbation at large scales decreases with scale separation. 
A simple two-modes model is derived that well describes the behaviors of energy concentration 
and growth-rates of various unstable flows. In the non-linear regime (at moderate values 
of $Re$) and in the presence of scale separation, the forcing scale and the largest scales 
of the system are found to be the most dominant energetically. 
\end{abstract}
%%%%%%%%%%%%%%%%%%%%%%%%%%%%%%%
\maketitle

%%%%%%%%%%%%%%%%%%%%%%%%%%%%%%%%%
\section{Introduction}        %%%
\label{sec:intro}             %%%
%%%%%%%%%%%%%%%%%%%%%%%%%%%%%%%%%

Hydrodynamic instabilities are responsible for the frequent encounter of turbulence 
in nature. Although instabilities are connected to the onset of turbulence and the 
generation of small scales, in many situation, instabilities are also responsible for 
the formation of large-scale structures. In such situations, flows of a given coherence 
length-scale are unstable to larger scale perturbations transferring energy to these scales.
A classical example of a large-scale instability is the $\alpha$-effect 
\cite{steenbeck_berechnung_1966,moffatt_field_1978} in magneto-hydrodynamic (MHD) 
flows to which the origin of large-scale planetary and solar magnetic field is attributed.
In $\alpha$-dynamo theory, small-scale helical flows self-organize to generate 
magnetic fields at the largest scale of the system.
%%%

%%%
While large-scale instabilities have been extensively studied for the dynamo problem, 
limited attention has been drawn to large-scale instabilities of the pure hydrodynamic case. 
Hence, most direct numeric simulations (DNS) and turbulence experiments are 
designed so that the energy injection scale $\ell$ is close to the domain size $L$. 
This allows to focus on the forward energy cascade and the formation of the 
Kolmogorov spectrum \cite{frisch_turbulence:_1995}.
Scales larger that the forcing scale, where no energy cascade is present, are expected 
\cite{frisch1985,Ganga_Fauve14} to reach a thermal equilibrium with a $k^2$ spectrum 
\cite{LEE:1952p4100,OrszagAnalytTheo,kraichnan73,krstulovicetal09}. Recent studies, 
using (hyper-viscous) simulations of turbulent flows randomly forced at intermediate 
scales \cite{Dallas}, have shown that the energy spectrum at large scales deviates 
from the thermal equilibrium prediction and forms a strong peak at the largest scale 
of the system. A possible explanation for this intriguing result is that a large-scale 
instability is present.
%%%

%%%
In pure hydrodynamic flows, the existence of large-scale instabilities has been 
known for some time. An asymptotic expansion based on scale separation %($\ell \ll L$)
was used in \cite{frisch_large-scale_1987,frisch_new_1988} to demonstrate the existence 
of a mechanism similar to the MHD $\alpha$-dynamo called the anisotropic-kinetic-alpha 
(\AKA{}) instability.
The \AKA{} instability is present in a certain class of non-parity-invariant, 
time-dependent and anisotropic flows. 
It appears for arbitrary small values of the Reynolds number and leads to a 
growth-rate $\sigma$ proportional to the wavenumber $q$ of the unstable 
mode: $\sigma\propto q$.
However, the necessary conditions for the presence of the \AKA{} instability are 
stricter than those of the $\alpha$-dynamo. 
Thus, most archetypal flows studied in the literature do not satisfy the \AKA{} conditions 
for instability. This, however, does not imply that the large scales are stable since other 
mechanisms may be present.
%%%

%%%
In the absence of an \AKA{}-effect higher-order terms in the large-scale expansion may lead to 
a so-called eddy-viscosity effect \cite{kraichnan_eddy_1976}.
This eddy-viscosity can be negative and thus produce a large-scale instability 
\cite{dubrulle_eddy_1991,wirth_eddy_1995}. 
The presence of a negative eddy-viscosity instability appears only above a critical 
value of the Reynolds number. It results in a weaker growth-rate than the \AKA{}-effect, 
proportional to the square of the wavenumber of the unstable mode $\sigma\propto q^2$.
Furthermore, the calculations of the eddy-viscosity coefficient can be much more difficult than 
those of the \AKA{} $\alpha$ coefficient. 
This difficulty originates on the order at which the Reynolds number enters the expansion
as we explain below. 
%%%

%%%
In the present paper, the Reynolds number is defined as $Re \equiv U_{rms} \ell / \nu$ 
where $U_{rms}$ is the root mean square value of the velocity and $\nu$ is the viscosity. 
Note that we have chosen to define the Reynolds number based on the energy injection scale $\ell$. 
An alternative choice would be to use the domain length scale $L$ which would lead to the 
large-scale Reynolds number that we will denote as $Re^L=UL/ \nu =(L/ \ell) Re$.
For the \AKA{} effect, the large-scale Reynolds number $Re^L$ is large, while 
the Reynolds number $Re$, based on the forcing scale $\ell$, is small. This allows to 
explicitly solve for the small-scale behavior and obtain analytic results. This is not possible for the 
eddy-viscosity calculation where there are two regimes to consider. Either the Reynolds numbers 
is small and the eddy-viscosity only provides a small correction to the regular viscosity, 
or the Reynolds numbers is large and the inversion of an advection operator is needed.
This last case can be obtained analytically only for very simple one dimensional shear flows 
\cite{dubrulle_eddy_1991,wirth_eddy_1995}.
%%%

\begin{figure}[!ht]
  \centering
  \includegraphics[width=\fwidth, trim= 750 0 0 0, clip=true]{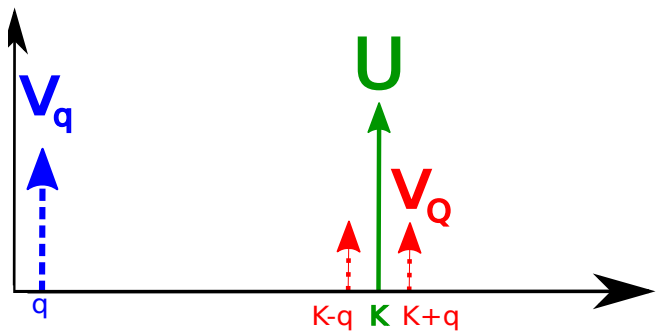}
  \caption{(Color online) Sketch of the three-modes model.
  $U$ represents the small-scale 
  driving flow of wavenumber $K$ (full arrow), $v_q$ is 
  the large-scale perturbation of wavevector $q$ (dashed arrow)
  and $v_Q$ is the small scale perturbation of wavevector $Q=K \pm q$ 
  (doted arrow).}
  \label{fig:3modmod}
\end{figure}

%%%
To illustrate the basic mechanisms involved in such multi-scale interactions, we depict in 
fig.~\ref{fig:3modmod} a toy model demonstrating the main ideas behind these instabilities. 
This toy model considers a driving flow, $\bm{U}$ at wavenumber ${\bf K}\sim1/\ell$, 
that couples to a small amplitude large-scale flow, $\bm{v}_{\bm{q}}$ at wavenumber 
${\bf q}\sim1/L$ with $\bf |q|\ll |K|$. The advection of $\bm{v}_{\bm{q}}$ by $\bm{U}$ and 
visa versa will then generate a secondary flow $\bm{v}_{\bm{Q}}$ at wavenumbers 
$\bm{Q}=\bm{K\pm q}$. This small-scale perturbation in turn couples to the driving flow 
and feeds back the large-scale flow. If this feedback is constructive enough to overcome 
viscous dissipation, it will amplify the large-scale flow and this process will lead to an 
exponential increase of $\bm{v}_{\bm{q}}$ and $\bm{v}_{\bm{Q}}$.
This toy model has most of the ingredients required for the instabilities to occur. 
%%%

%%%
%%%%%%%%%%%%%%%%%%%%%%%%%%%%%%%%%
%%%%%        FIG 1       %%%%%%%%
%%%%%%%%%%%%%%%%%%%%%%%%%%%%%%%%%
%%%%%%%%%%%%%%%%%%%%%%%%%%%%%%%%%
%%%

%%%
In order to study large-scale instabilities, they must be isolated from other small-scale 
competing instabilities that might coexist. This can be achieved by using Floquet theory 
\cite{floquet_sur_1883} (also referred as Bloch theory in quantum mechanics \cite{ashcroft_solid_1976}). 
Indeed, Floquet theory can track modes with large and small spatial periodicity separately. 
In what follows, we use direct numerical simulations (DNS) in the Floquet framework 
to study different flows, either in the presence of the \AKA{} effect using the flow introduced 
in \cite{frisch_large-scale_1987} or in the absence of \AKA{} effect using the \dombre{} flow 
(A=B=C) \cite{dombre_chaotic_1986} and the Roberts flow \cite{roberts_spatially_1970}. 
Our study extends to values of $Re$ and $\ell/L$ beyond the range of validity of the 
asymptotic expansions. Finally, we compare the results of Floquet 
DNS to those of full Navier-Stokes DNS.
%%%

%%%
%%%%%%%%%%%%%%%%
\section{Methods}              %%
\label{sec:methods}           %%
%%%%%%%%%%%%%%%%
\subsection{Navier-Stokes}
\label{subsec:navierStokes}
%%%

%%%
Our starting point is the incompressible Navier-Stokes equation in the periodic $[0, 2\pi L]^3$-cube:
\begin{align}
	\dt \Ugen 
	= \Ugen \X \roT \Ugen - \grad \Pgen
	+ \visco \Laplace \Ugen +\Fgen \, ,
%	\, \diV \Ugen =0 \, ,
 \label{eq:fullNS}
\end{align}
with $\diV \Ugen =0$ and 
where $\Ugen$, $\Fgen$, $\Pgen$ and $\nu$ denote the velocity field,
the forcing field, the generalized pressure field and the viscosity coefficient, respectively.
The geometry imposes that all fields be $2\pi L$-periodic. We further assume that the 
forcing has a shorter spatial period $2\pi\ell$ with $L/\ell$ an arbitrary large integer. 
We denote the wavenumber of this periodic forcing as $\bf K$, with $K=|{\bf K}|=1/\ell$ for 
the flows examined. If the initial conditions of $\bf V$ satisfies the same periodicity as 
$\Fgen$ then this periodicity will be preserved by the solutions of the Navier-Stokes
and corresponds to the preservation of the discrete symmetries 
$x\to x+2\pi\ell$, $y\to y+2\pi\ell$ and $z\to z+2\pi \ell$.
However, these solutions can be unstable to arbitrary small perturbations that break
this symmetry and grow exponentially. To investigate the stability of the periodic solutions, 
we decompose the velocity and pressure field in a driving flow and a perturbation component:
\begin{align}
	\Ugen= \Ulam+\vlin 
		\quad , \quad
	\Pgen = \Pgen_{\Ulam} + \Pgen_{\vlin}
\end{align}
where $\Ulam$ denotes the driving flow that has the same periodicity as 
the forcing $2\pi \ell$ and $\vlin$ is the velocity perturbation. 
The linear stability analysis amounts to determining the evolution of small 
amplitude perturbations so that only the first order terms in $\vlin$ are kept. 
The evolution equation of the driving flow is thus:
\begin{align}
	\dt \Ulam = \Ulam \X \roT \Ulam - \grad \Pgen_{\Ulam} 
	+ \visco \Laplace \Ulam + \Fgen \, .
 \label{eq:linNS:K}
\end{align}
The remaining terms give the linearized Navier-Stokes equation for the perturbation:
\begin{align}
	\dt \vlin
	= \Ulam \X \roT \vlin+& \vlin \X \roT \Ulam 
	- \grad \plin_{\vlin} + \visco \Laplace \vlin \, ,
%	\, \diV\vlin=0 
%	\\ \text{with} \quad \diV\vlin&=0 \, . 
 \label{eq:linNS:rK}
\end{align}
The two pressure terms enforce the incompressibility conditions $\diV\Ulam=0$ and $\diV\vlin=0$.
The $\bf U$ flow is not necessarily a laminar flow (but respects $2\pi \ell$ periodicity). 
In general, the linear perturbation $\vlin$ does not only consist of modes that break the 
periodicity of the forcing. Linear unstable modes respecting the periodicity may also exist: 
they correspond to small-scale instabilities. We show how these modes can be distinguished 
from periodicity-breaking large-scale modes in the following section devoted to Floquet analysis.
%%%

%%%
%%%%%%%%%%%%%%%%%%%%%%%%%%%%%%%%
\subsection{Floquet Analysis} %%
\label{subsec:FLASH}               %%
%%%%%%%%%%%%%%%%%%%%%%%%%%%%%%%%
%%%

%%%
Studying large-scale flow perturbations with a code that solves 
the full Navier-Stokes equation requires considerable computational power
as resolution of all scales from domain size $L$ to 
the smallest viscous scales $\ell_\nu\ll \ell$ must be achieved.
This is particularly difficult in our case where scale separation $\ell\ll L$ is required.
In order to overcome this limitation, we adopt the Floquet framework \cite{floquet_sur_1883}. 
In Floquet theory, the velocity perturbation can be decomposed into modes 
that are expressed as the product of a complex harmonic wave, $e^{i \qvec \cdot \bf r}$,
multiplied by a periodic vector field $\vfloq(\rvec,t)$ with the same periodicity $2\pi \ell$ 
as that of the driving flow:
\begin{align}
 \vlin(\rvec,t) = \vfloq(\rvec,t)e^{\imath \qvec \cdot \rvec } + c.c. \, ,
\end{align} 
 and similar for the pressure,
\begin{align}
 \plin_{\vlin}(\rvec,t) = \pfloq(\rvec,t)e^{\imath \qvec \cdot \rvec } + c.c. \, ,
\end{align}
where \textit{c.c.} denotes the complex conjugate of the previous term.
%%%

%%%
Perturbations whose values of $\bf q$ are such that at least one component is 
not an integer multiple of $1/\ell$, break the periodicity of the driving flow.
The perturbation field $\vlin$ then involves all Fourier 
wavenumbers of the type $\bf Q=q+k$, where $\bf k$ is a wavevector corresponding to
the $2\pi\ell$-periodic space dependence of $\vfloq$.
We restrict the study to values of $q=|\bf q|$ satisfying $0 < q \le K$.
For finite domain sizes $\bf q$ is a discrete vector with ${q}\ge 1/L$, while for infinite 
domain sizes $\qvec$ can take any arbitrarily small value.
In the limit ${ q/K}\ll 1$ the perturbation involves scales much larger than $\ell$. 
Therefore, scale separation is achieved without solving intermediate scales as would 
be required if the full Navier-Stokes equations were used. Furthermore, this framework 
has the advantage of isolating perturbations that break the forcing periodicity 
($\qvec \ell \notin \mathbb{Z}^3$), from other small-scale unstable modes with the 
same periodicity ($\qvec \ell \in \mathbb{Z}^3$) that might also exist in the system. 
%%%

%%%
A drawback of the Floquet decomposition is that some operators 
have somewhat more complicated expressions than in the
simple periodic case. For instance, taking 
a derivative requires to take into account the variations of both the 
harmonic and the amplitude. Separating the amplitude in its 
real and imaginary parts $\vfloq(\rvec,t)=\vfloq^{r} + \imath \vfloq^{i}$,
we obtain
\begin{align}
 \partial_x \vlin % = (iq_x + \partial_x) (\vfloq^{r} + \imath \vfloq^{i}) 
 = \left[ \partial_x \vfloq^{r} - q_x \vfloq^{i} + 
 \imath (q_x \vfloq^{r} +\partial_x \vfloq^{i}) \right]
 e^{\imath \qvec \cdot \rvec } + c.c. \,,
 \label{eq:der:Floq}
\end{align}
where $\partial_x$ denotes the $x$-derivative and 
$q_x$ denotes the $x$-component of the $\qvec$ wavevector.
%%%

%%%
Using eq.~\eqref{eq:linNS:rK} and \eqref{eq:der:Floq}, the linearized 
Navier-Stokes equation can be written as a set of $3+1$ 
complex scalar equations:
\begin{align}
 \nonumber
 \partial_t \vfloq =& (\roT\Ulam) \times \vfloq + 
 (\imath \qvec \X \vfloq + \roT \vfloq ) \times \Ulam
 \\ & 
 - (\imath \qvec + \grad) \pfloq + \visco ( -\qvec^2 + \Delta) \vfloq \, ,
 \label{eq:ns:Flq} \\
 \text{with} \quad & \imath \qvec \cdot \vfloq + \diV \vfloq=0 \, .
 \label{eq:incomp:Flq}
\end{align}
%%%

%%%
We use standard pseudo-spectral methods to solve this system of equations
in the $2\pi \ell$-periodic cube.
The complex velocity field $\vfloq$ is decomposed in Fourier space where 
derivatives are reduced to a multiplication by $\imath \kvec$, 
where $\kvec$ is the Fourier wavevector. Multiplicative term are computed
in real space. These methods have been implemented in the: 
Floquet Linear Analysis for Spectral Hydrodynamics (FLASH) code 
and details are given in appx.~\ref{sec:apx:FLASH}.
%%%

%%%
In order to find the growth-rate of the most unstable mode, we integrate 
eq.~\eqref{eq:ns:Flq},\eqref{eq:incomp:Flq}, for a time long enough for a clear 
exponential behaviour to be observed. The growth-rate of this most 
unstable mode can then be measured by linear fitting. Note that this 
process only leads to the measurement of the fastest growing mode.
%%%

%%%
%%%%%%%%%%%%%%%%%%%%%%%%%%%%%%%%
\subsection{Three-modes model} %
\label{subsec:3mm}                        %
%%%%%%%%%%%%%%%%%%%%%%%%%%%%%%%%
%%%

%%%
Although the Floquet framework is very convenient to solve equations
numerically, it does not easily yield analytic results. 
Rigorous results must be based on asymptotic expansions and can only be 
derived in the limit of small \Rn{} or for simple shear layers 
\cite{dubrulle_eddy_1991,wirth_eddy_1995}. 
To obtain a basic understanding of the processes involved, we will 
use the idea represented in the toy model of fig.~\ref{fig:3modmod}. 
This model also has the major advantage of using a formalism 
that can easily be related to the physical aspect of the problem.
%%%

\begin{figure}[!ht]
  \centering
  \includegraphics[width=\fwidth]{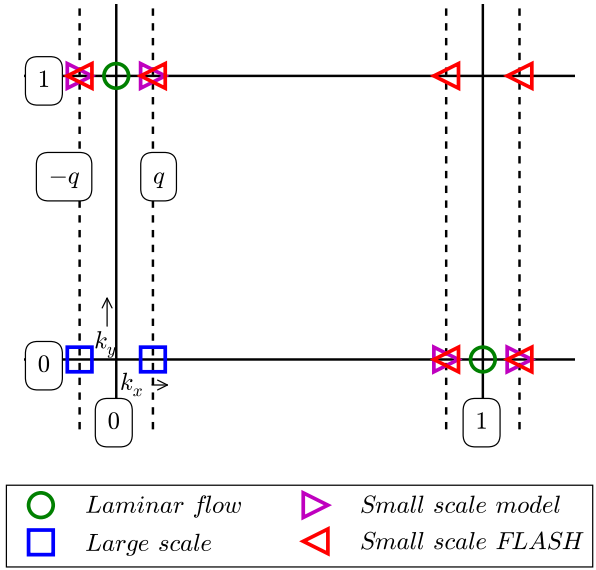}
  \caption{Fourier modes of the Floquet decomposition used 
  in the FLASH code and the three-modes model.}
  \label{fig:floquetSpec}
\end{figure}

%%%
In our derivation, we only consider the evolution of the two most intense 
modes of the perturbation and of the driving flow. The velocity perturbation 
is thus decomposed as a series of velocity fields of different modes: 
\begin{align}
	\vlin(\rvec, t) &= \pmv(\rvec, t) + \pmV(\rvec, t) + \rmv(\rvec,t) \, ,
	\\
	\pmv(\rvec, t) &= \vfloq(\qvec, t) e^{\imath \qvec \rvec } + c.c. \, , 
	\\
	\pmV(\rvec, t) &= \sum_{\knrm=1} \vfloq(\qvec,\kvec, t) 
	e^{\imath ( \qvec \cdot \rvec + \kvec \cdot \rvec ) } + c.c. \, ,
	\\
	\rmv(\rvec, t) &= \sum_{\knrm > 1} \vfloq(\qvec,\kvec, t) 
	e^{\imath ( \qvec \cdot \rvec + \kvec \cdot \rvec ) } + c.c. \, , 
\end{align} 
where $\qvec$ denotes the wavenumber of the large-scale modes and $\Qvec$ 
denotes the modes directly coupled to $\qvec$ via the driving flow, since $K=1$. 
At wavenumber $\qvec$, the linearized Navier-Stokes equation can be rewritten as:
\begin{align}
 \dt \pmv = \Ulam\X\roT\pmV +\pmV \X \roT \Ulam
 - \grad \pmp + \visco \Laplace \pmv \,.
 \label{eq:linNS:q}
\end{align}
Assuming that the coupling with the truncated velocity, $\rmv$, is negligible 
with respect to the coupling with the large-scale velocity, $\pmv$, the linearized 
equation at $\Qvec$ reads:
\begin{align}
 \dt \pmV = \Ulam \X \roT\pmv + \pmv\X \roT\Ulam
 - \grad \pmP + \visco \Laplace \pmV \, ,
 \label{eq:linNS:Kpmq}
\end{align}
where $\pmp$ and $\pmP$ denote the pressure enforcing the 
incompressible conditions: $\diV \pmv=0$ and $\diV \pmV=0$, respectively. 
The modes are represented in fig.~\ref{fig:floquetSpec}.
%%%%%%%%%%%%%%%%%%%%%%%%%%%%%%%%%
%%%%%        FIG 2       %%%%%%%%
%%%%%%%%%%%%%%%%%%%%%%%%%%%%%%%%%
%%%%%%%%%%%%%%%%%%%%%%%%%%%%%%%%%
%%%

%%%
The derivation is restricted to stationary positive helical driving flows, satisfying:
$
	\Uhel ( \rvec ) = K^{-1} \roT \Uhel( \rvec ) \,.
$
The problem can then be solved by making use of the vorticity fields:
\begin{align}
	\pmomega = \roT \pmv
	\quad \text{and} \quad
	\pmOmega = \roT \pmV \,,
	\label{eq:pmOmegas}
\end{align}
and the adiabatic approximation: $\partial_t \pmV \ll \nu \Laplace\pmV $.
The system of equations of the three-modes model is thus:
\begin{align}
	\visco \Laplace \pmOmega &= 
	- \roT \left [ \Uhel \times ( \pmomega - K \pmv ) \right] \, , 
	\\
	\dt \pmomega &= \roT \left [ \Uhel \X (\pmOmega - K \pmV )\right ] + 
	\visco \Laplace \pmomega \, .
\end{align}
The greatest eigenvalue of the system, $\sigma$, gives the growth-rate of the 
perturbation. The growth-rate can be derived analytically for an \ABC{} 
large-scale flow:
\begin{align}
 U^{ABC}_x &= C \sin( K z ) + B \cos( K y ) \, ,
  \\
 U^{ABC}_y &= A \sin( K x ) + C \cos( K z ) \, ,
  \\
 U^{ABC}_z &= B \sin( K y ) + A \cos( K y ) \, .
 \label{eq:ABC:flow}
\end{align}
For \abc{} flows (\lamABC{}), one finds:
\begin{align}
 \sigma = \beta q^2 - \nu q^2
  \quad &\text{with} \quad
  \beta = \bc \rek^2 \nu \, , 
  \label{eq:sigABC0}
    \\
  \bc = \frac{1-\lambda^2}{4+2\lambda^2} 
  \quad &\text{and} \quad 
  \rek = \tfrac{U}{K \nu} \, ,
 \label{eq:sigABC}
\end{align}
where \rek{} denotes the small-scale \Rn{} defined using the driving flow. 
The fastest growing mode is found to be fully helical.
%%%

%%%
This simple model indicates that some driving flows, not satisfying the hypotheses 
of the \AKA{}-effect, described in \cite{frisch_large-scale_1987}, can generate 
a negative eddy-viscosity instability satisfying $\sigma \propto q^2$. 
The largest growth-rate is obtained for $\lambda=0$ while no $q^2$ instability is 
predicted for $\lambda=1$. For $\lambda \ne 1$ the flow becomes unstable when the $\beta$ 
term can overcome the viscosity $\beta>\nu$. This happens when $Re$ is above a critical 
value: $\rec= \bc^{-1/2}$.
%%%

%%%
%%%%%%%%%%%%%%%%%
%%%%%%%%%%%%%%%%%
\section{Results}            %%%%
\label{sec:results}          %%%% 
%%%%%%%%%%%%%%%%%
%%%%%%%%%%%%%%%%%
%%%

%%%
%%%%%%%%%%%%%%%%
\subsection{\AKA{}}           %%
\label{subsec:aka}              %%
%%%%%%%%%%%%%%%%
%%%

%%%
We begin by examining a flow that satisfies the conditions for an \AKA{} instability.
Such a flow was proposed in \cite{frisch_large-scale_1987} (from now on $Fr87$)
and is given by:
\begin{align}
  U^{Fr87}_x &= U_0
  \cos \left( K y + \visco K^2 t \right) \, , % - \frac{\pi}{4} 
  \nonumber
  \\
  U^{Fr87}_y &= U_0
  \sin \left( K x - \visco K^2 t \right) \, , %+ \frac{\pi}{4} 
  \label{eq:Fr87:flow}
  \\
  U^{Fr87}_z &= U^{Fr87}_x + U^{Fr87}_y \, . 
  \nonumber
\end{align} 
The growth-rate of large-scale unstable modes can be calculated in the small 
Reynolds number limit and is given by:
\begin{align}
 \sigma = \alpha q - \nu q^2 \, ,
%  \quad \text{with} \quad
% \alpha = a \rek U_0 
% \quad \text{and} \quad
% 	\ac = \frac{1}{2} \, .
 \label{eq:Fr87:alpha}
\end{align} 
with $\alpha=a Re U_0$ and $a=\frac{1}{2}$.
The fastest growing mode has negative helicity and \qvec{} along the $z$-direction.
%%%

\begin{figure}[!ht]
  \centering
  \includegraphics[width=\fwidth , trim= 5 10 5 5, clip=true]{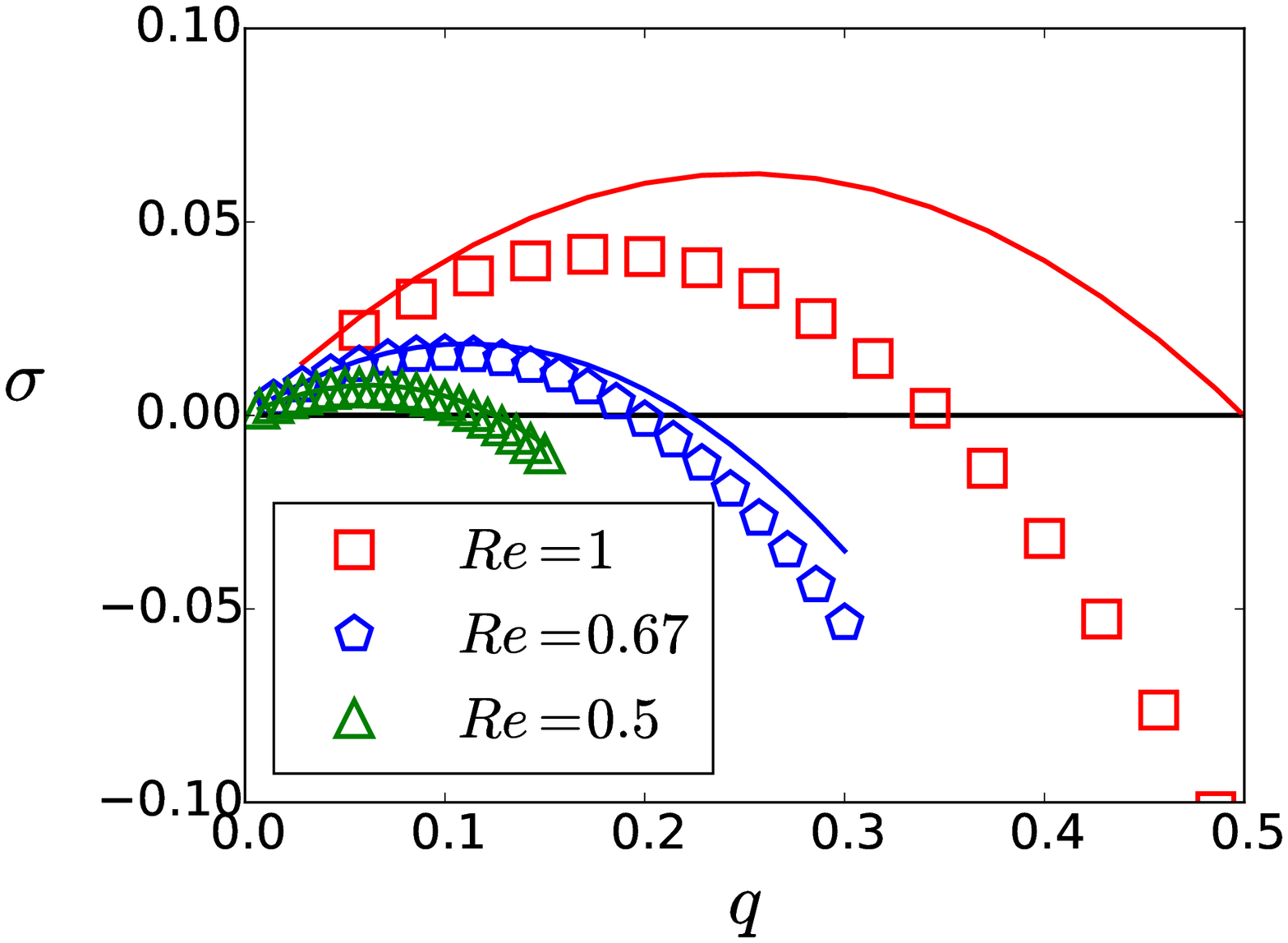}
  \caption{Growth-rate \textit{vs.} Floquet wavenumber, $\sigma(q)$, at different $\rek$ 
  plotted in log-log scale for a $Fr87 $flow, eq.~\eqref{eq:Fr87:flow}.}
  \label{fig:sigVqVreAKA_lin}
\end{figure}

\begin{figure}[!ht]
  \centering
  \includegraphics[width=\fwidth,trim= 5 10 5 5, clip=true]{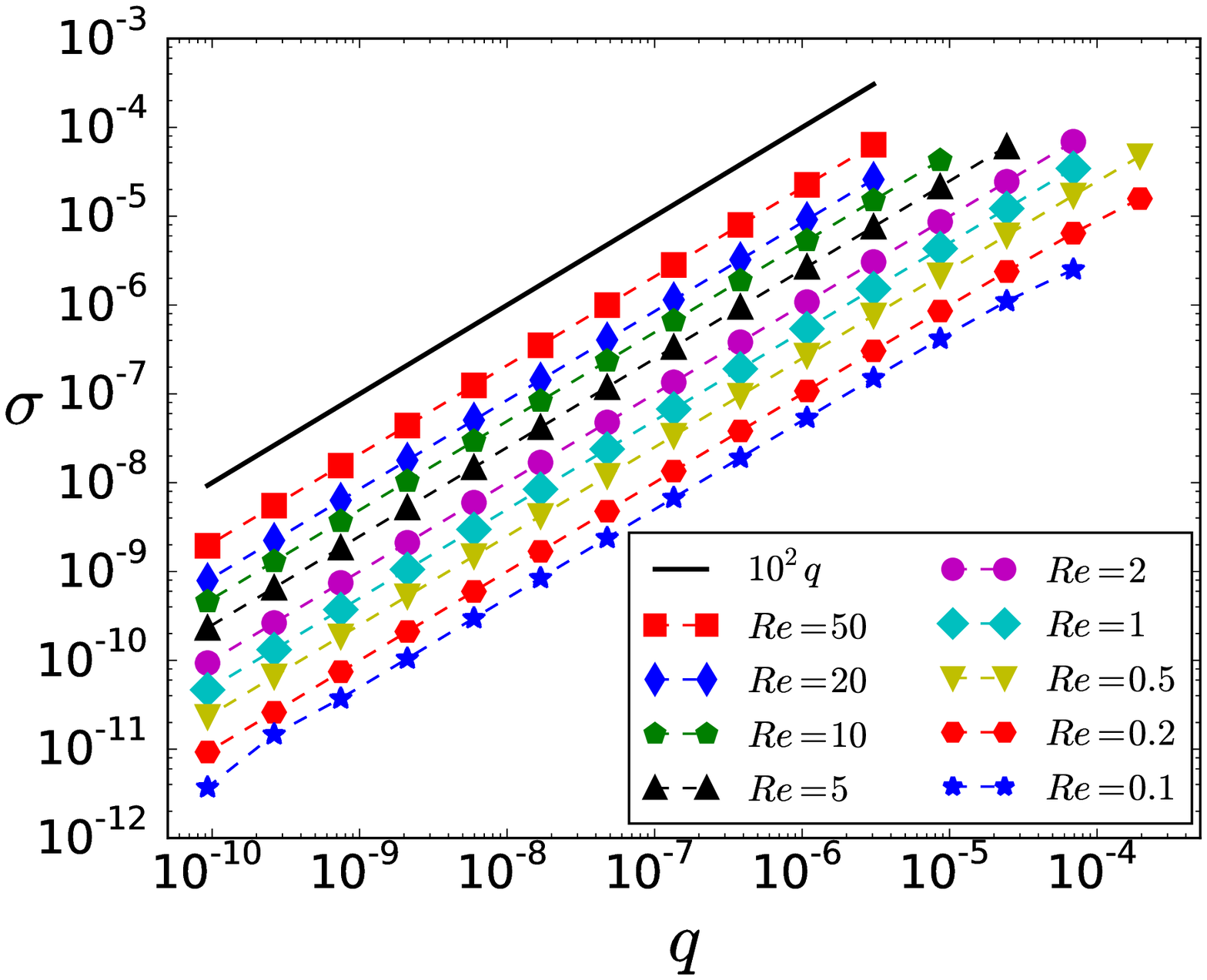}
  \caption{Growth-rate \textit{vs.} Floquet wavenumber, $\sigma(q)$, 
  at different $\rek$ plotted in log-log scale for a $Fr87$ flow, 
  eq.~\eqref{eq:Fr87:flow}.}
  \label{fig:sigVqVreAKA}
\end{figure}

\begin{figure}[!htb]
    \centering
    \begin{tikzpicture}
        \node[anchor=south west,inner sep=0] (image) at (0,0) 
        {\includegraphics[width=\fwidth,trim= 0 25 35 15, clip=true]{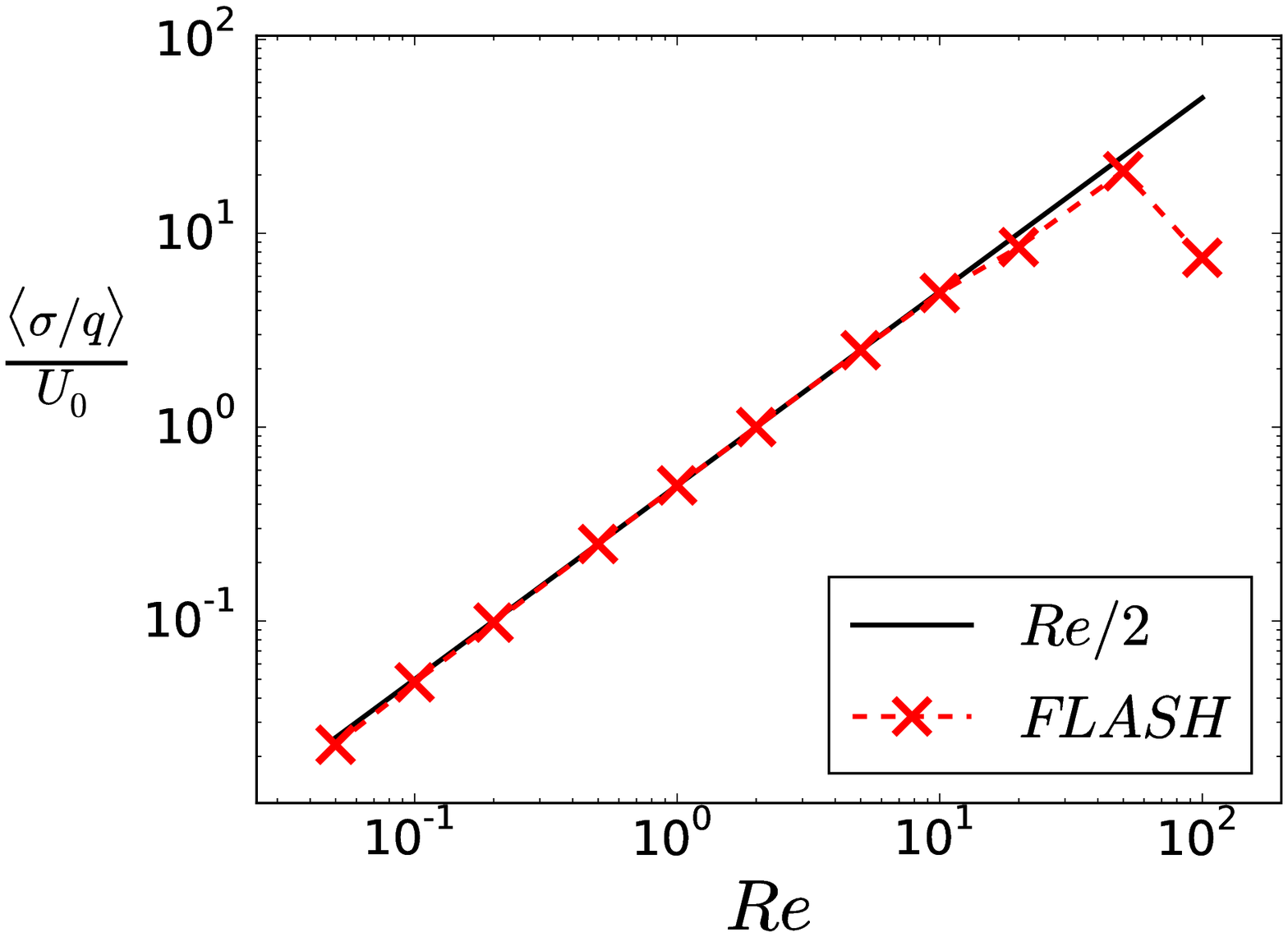}};
        \begin{scope}[x={(image.south east)},y={(image.north west)}]
%            \draw[help lines,xstep=.05,ystep=.05] (0,0) grid (1,1);
%            \foreach \x in {0,1,...,9} { \node [anchor=north] at (\x/10,0) {0.\x}; }
%            \foreach \y in {0,1,...,9} { \node [anchor=east] at (0,\y/10) {0.\y}; }
            \node[anchor=south west,inner sep=0] (image) at (0.2,0.6) 
            {\includegraphics[width=3.7cm,trim= 0 25 35 15, clip=true]{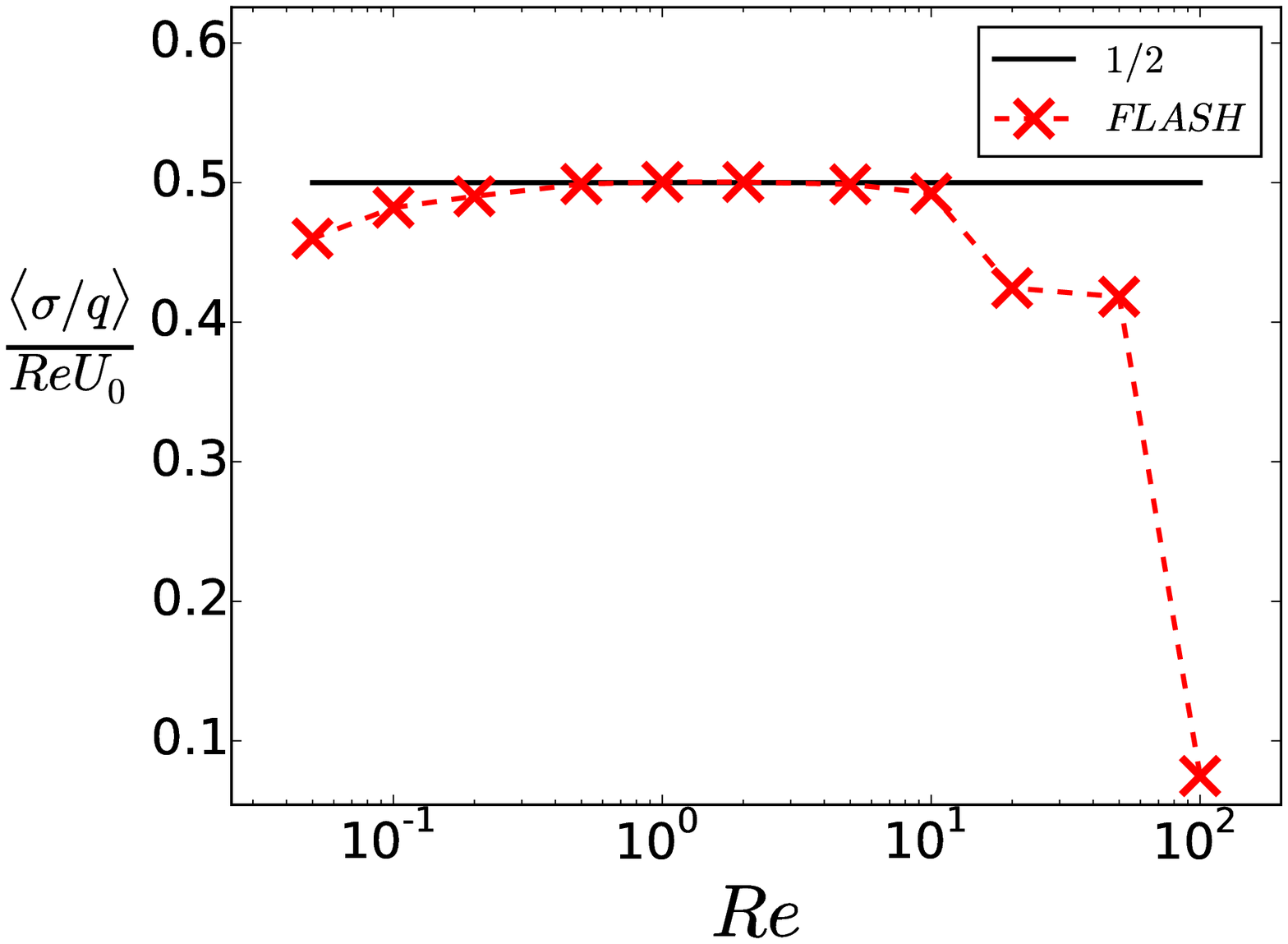}};
        \end{scope}
    \end{tikzpicture}
	\caption{\alphaMes{}~coefficient \textit{vs.} \Rn{}, % \alphaMes{}$(\rek)$, 
	plotted in log-log scale for an instability generated by a $Fr87$ flow, eq.~\eqref{eq:Fr87:flow}. 
	In insert \alphaReMes{} \textit{vs.} \Rn{} plotted in lin-log scale.}
    \label{fig:alphaVreAKA}
\end{figure}

%%%
Setting \qvec{} along the $z$-direction, we integrated eq.~\eqref{eq:incomp:Flq})
numerically and measured the growth-rate $\sigma$. Fig.~\ref{fig:sigVqVreAKA_lin} 
displays the growth-rate of the most unstable mode as a function of the wavenumber 
amplitude $q=|\qvec |$ for three different values of $Re$ measured by the Floquet code 
and compared to the theoretical prediction. 
%%%%%%%%%%%%%%%%%%%%%%%%%%%%%%%%%
%%%%%        FIG 3       %%%%%%%%
%%%%%%%%%%%%%%%%%%%%%%%%%%%%%%%%%
The agreement is good for small values of $q$ and for small values of $Re$ where 
the asymptotic limit is valid. For $q$ small enough, the flow is unstable and
satisfies $\sigma \propto q$. 
Fig.~\ref{fig:sigVqVreAKA} shows in log-log scale the growth-rate of the 
perturbation as a function of $q$ for different Reynolds numbers. 
The solid line in the graph indicates the $\sigma \propto q$ scaling which 
is satisfied for all $Re$.
%%%%%%%%%%%%%%%%%%%%%%%%%%%%%%%%%
%%%%%        FIG 4       %%%%%%%%
%%%%%%%%%%%%%%%%%%%%%%%%%%%%%%%%%
%%%%%%%%%%%%%%%%%%%%%%%%%%%%%%%%%
In fig.~\ref{fig:alphaVreAKA}, 
we compare the theoretical and numerically calculated prefactor 
$a$ of the $\alpha$~coefficient. This coefficient increases linearly 
with $Re$ and is seen to be in good agreement with the theoretical 
prediction up to $Re\simeq 10$. For larger values of $Re$, $a$ 
deviates from the linear prediction and saturates. 
%%%

%%%
%%%%%%%%%%%%%%%%%%%%%%%%%%%%%%%%%
%%%%%        FIG 5       %%%%%%%%
%%%%%%%%%%%%%%%%%%%%%%%%%%%%%%%%%
%%%%%%%%%%%%%%%%%%%%
%%%

\begin{figure}[!ht]
  \centering
  \includegraphics[width=\fwidth,trim= 5 10 5 5, clip=true]{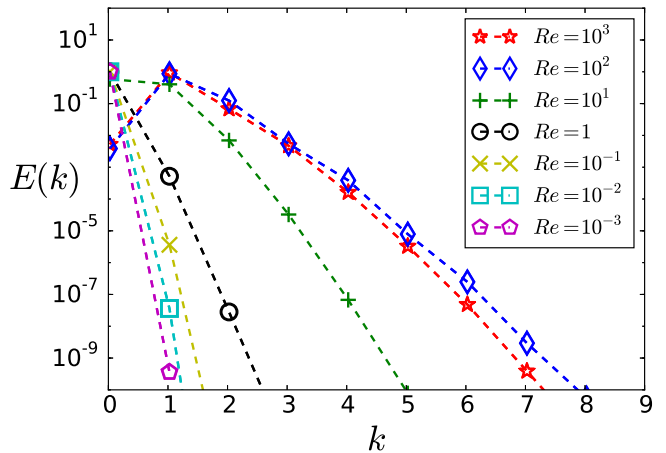}
  \caption{Spectrum of the Floquet perturbation, $E(k)$, for 
	different small-scale Reynolds numbers, $\rek$, with 
	$\qvec = (0; 0; 0.025)$ generated by a $Fr87$ flow, eq.~\eqref{eq:Fr87:flow}.}
  \label{fig:specAKA}
\end{figure}

%%%
A positive growth-rate for a small $q$ mode does not guarantee 
the dominance of large scales. We should also consider what fraction of the 
perturbation energy is concentrated in the large scales. Fig.~\ref{fig:specAKA} 
shows the energy spectra for different Reynolds numbers. The energy spectrum 
for the complex Floquet field $\vfloq{}$ is defined as:
$E(k) = \sum_{k-\frac{1}{2}\le {\vert \bf k \vert} \leq k+\frac{1}{2}} \vert \vfloq{} \vert^2 $ 
with $E(k=0)$ the energy at large scales $1/q$. While at small Reynolds numbers, the 
smallest wavenumber $k=0$ dominates, as the Reynolds number increases,
more energy is concentrated in the wavenumber of the driving flow $k=1$. 
%%%

%%%
%%%%%%%%%%%%%%%%
%%%%%        FIG 6         %%%
%%%%%%%%%%%%%%%%
\begin{figure}[!ht]
  \centering
  \includegraphics[width=\fwidth,trim= 5 8 5 5, clip=true]{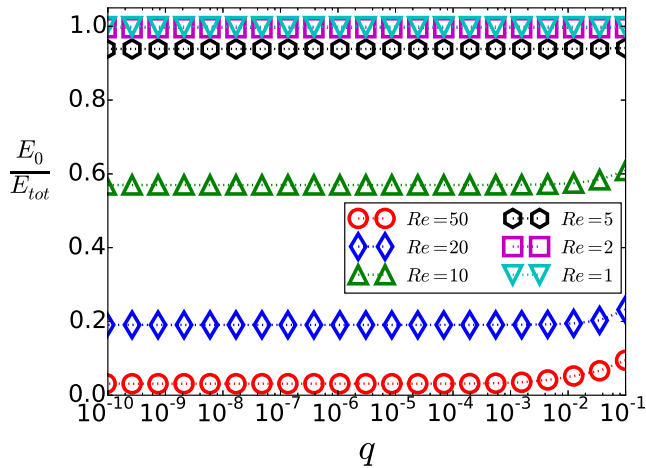}
  \caption{Growth-rate \textit{vs.} Floquet wavenumber, $\sigma(q)$, at different $\rek$ 
  plotted in log-log scale for a $Fr87$ flow, eq.~\eqref{eq:Fr87:flow}.}
  \label{fig:specAKAfracq}
\end{figure}

\begin{figure}[!ht]
  \centering
  \includegraphics[width=\fwidth,trim= 5 10 35 25, clip=true]{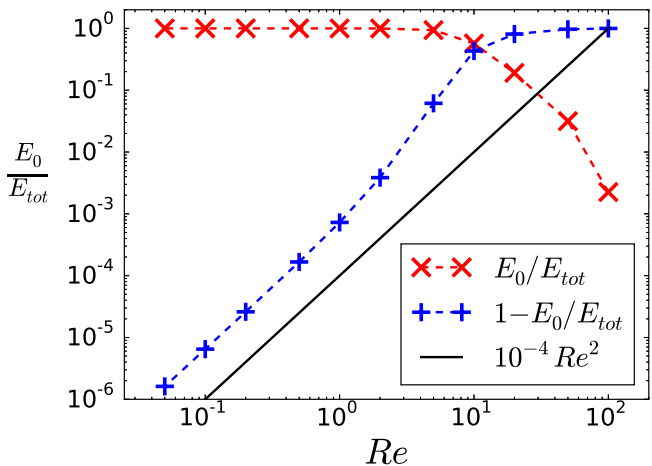}
  \caption{Energy ratio \textit{vs.} Reynolds number, $E_0$, 
  plotted in log-log scale for a $Fr87$ flow, eq.~\eqref{eq:Fr87:flow}.}
  \label{fig:specAKAfracRe}
\end{figure}

%%%%%%%%%%%%%%%%
To quantify this behavior, we plot in fig.~\ref{fig:specAKAfracq} the 
fraction of the energy in the zero mode $E_0 = E(0)$ divided by the total 
energy of the perturbation $E_{tot}=\sum_{k=0}^{\infty}E(k)$, as a function 
of the wavenumber $q$ for different values of $Re$. In the small $q$ limit, 
this ratio reaches an asymptote that depends on the Reynolds number. 
This asymptotic value is shown as a function of the $Re$ in 
fig.~\ref{fig:specAKAfracRe}. The small-scale energy ($E_{tot}-E_0$) 
is then shown to follow a power law $1-\frac{E_{0}}{E_{tot}}\propto Re^{2}$ 
for small values of $Re$.
%%%%%%%%%%%%%%%%
Therefore, for the \AKA{} instability, at small $Re$, the energy 
is concentrated in the large scales, whereas, at large 
$Re$, the most unstable mode has a small projection in the large scales.
%%%%%%%%%%%%%%%%
%%%%%        FIG 7         %%%
%%%%%%%%%%%%%%%%
%%%%%%%%%%%%%%%
%%%%%%%%%%%%%%%
%%%%%        FIG 8        %%
%%%%%%%%%%%%%%%
%%%

%%%
%%%%%%%%%%%%%%%%%%%%%%%%%%%%%%%%%%%%%%%%%%
\subsection{Roberts flow: $\lambda=0$ } %%
\begin{figure}[!ht]
  \centering
  \includegraphics[width=\fwidth,trim= 5 5 5 5, clip=true]{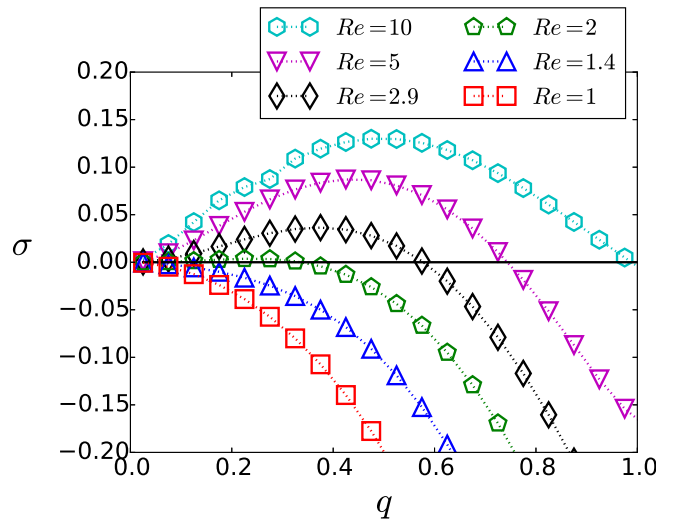}
  \caption{Growth-rate $\sigma$ \textit{vs.} Floquet wavenumber $q$, for different $\rek$ 
   for the Roberts flow.}
  \label{fig:linRob}
\end{figure}

\label{subsec:rob}                      %%
%%%%%%%%%%%%%%%%%%%%%%%%%%%%%%%%%%%%%%%%%%
We now investigate non-\AKA{}-unstable flows. We 
consider the family of the \ABC{} flow, for which we expect large-scale 
instabilities of the form given in eq.~\eqref{eq:sigABC}. The three-modes model 
predicts that from the family of \ABC{} flows the most unstable is the 
\rob{} flow that is commonly referred to as the Roberts flow in the literature
\cite{roberts_spatially_1970}. The model predicts a positive growth-rate when $Re>2$. 
Fig.~\ref{fig:linRob} shows the growth-rate $\sigma$ as a function of $q$ for various 
Reynolds numbers calculated using the Floquet code. 
For small values of the Reynolds number all modes $q$ have negative growth-rate.
Above a critical value $\rec\simeq 2$ unstable modes appear at small values of $q$
in agreement with the model predictions.
%%%

%%%
%%%%%%%%%%%%%%%
%%%%%        FIG 9         %%
%%%%%%%%%%%%%%%
%%%%%%%%%%%%%%%
%%%

\begin{figure}[!ht]
  \centering
  \includegraphics[width=\fwidth,trim= 5 5 5 5, clip=true]{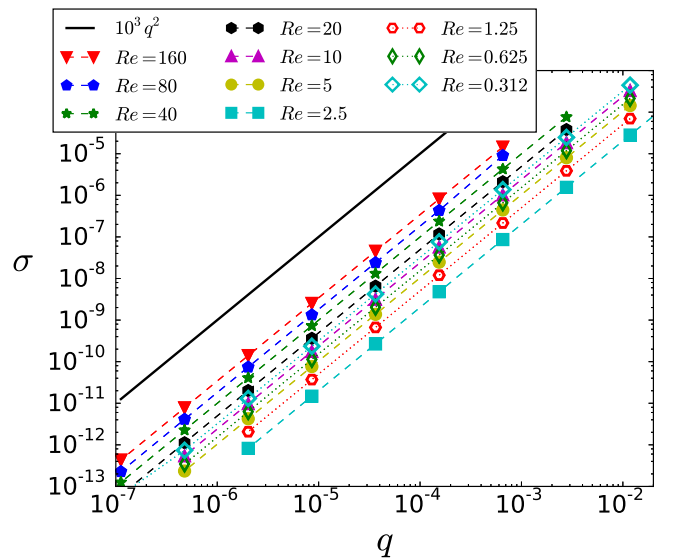}
  \caption{Growth-rate \textit{vs.} Floquet wavenumber, $\sigma(q)$, for different $\rek$ 
  plotted in log-log scale for a Roberts flow. The full markers with dashes represent
  the value of positive growth-rates whereas the empty markers with dots represent
  the absolute value of negative growth-rates.}
  \label{fig:sigVqVreRob}
\end{figure}

%%%
To investigate the behavior of the instability for small values of $q$
we plot in Fig.~\ref{fig:sigVqVreRob}
the absolute value of the growth-rate as a function of $q$, in a logarithmic 
scale, for \Rn{} ranging from $0.312$ to $160$. Dashed lines indicate positive 
growth-rates while dotted lines indicate negative growth-rates. The solid 
black line indicates the $\sigma \propto q^2$ scaling followed by all curves.
Therefore, the scaling predicted by the model (eq.~\eqref{eq:sigABC0},\eqref{eq:sigABC}) 
is verified. We will refer to the instabilities that follow this scaling $\sigma \propto q^2$ 
as negative eddy-viscosity instabilities.
%%%%%%%%%%%%%%%%%%%%%%%%%%%%
%%%%%        FIG 10       %%
%%%%%%%%%%%%%%%%%%%%%%%%%%%%
%%%

\begin{figure}[!ht]
  \centering
  \includegraphics[width=\fwidth,trim= 5 10 5 5, clip=true]{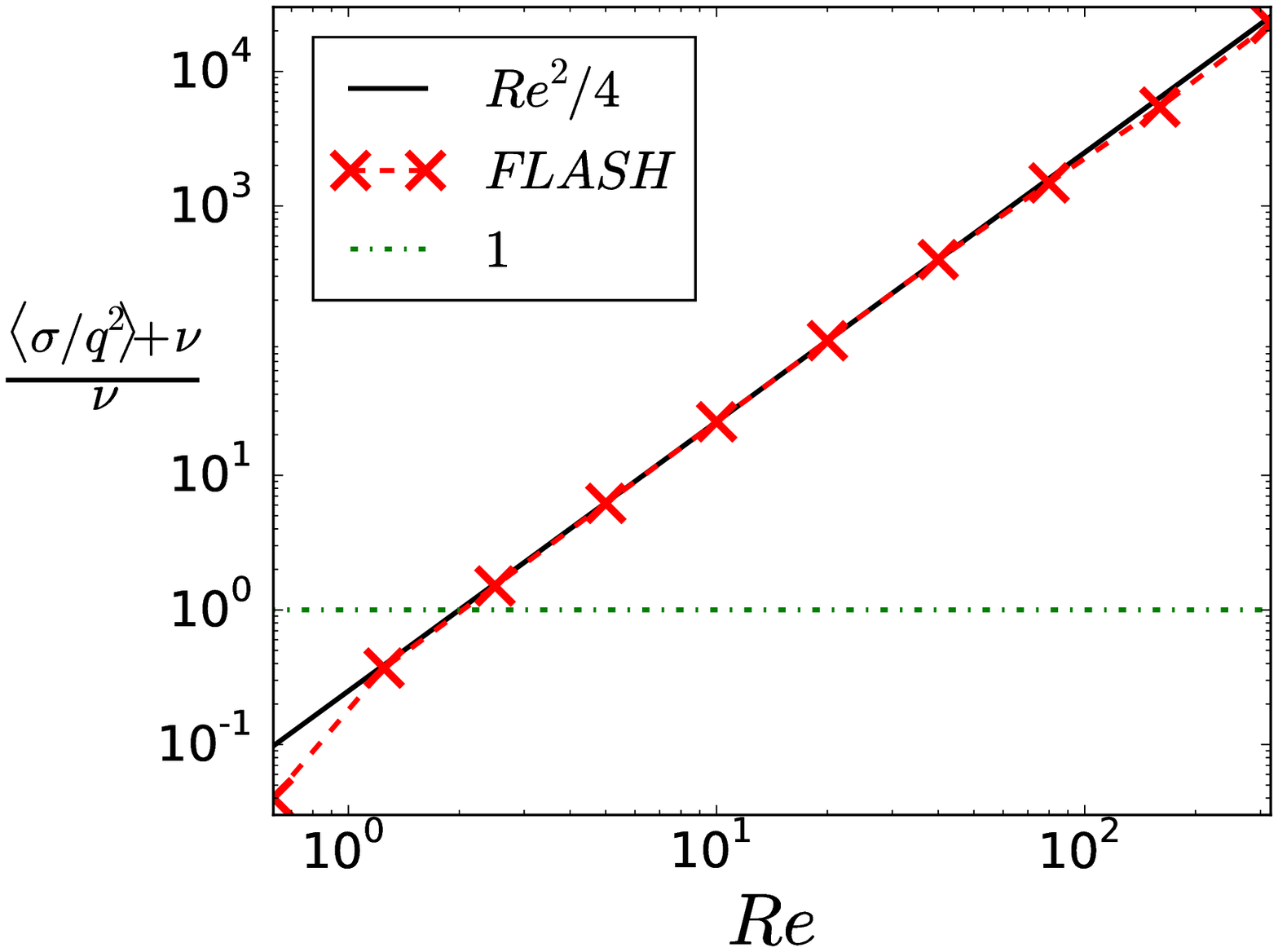}
  \caption{\betaMes{} \textit{vs.} of \Rn{},
  plotted in lin-log scale for the Roberts flow. 
  }
  \label{fig:betaVreRob}
\end{figure}

%%%
To further test the model predictions we measure the proportionality coefficient 
for the $q^2$ power law obtained from the Floquet code.
Fig.~\ref{fig:betaVreRob} compares the \bc{}~coefficient predicted by the three-modes 
model with the results of the Floquet code.
The figure shows $(\langle \sigma/q^2\rangle + \nu)/\nu$ measured from the data for 
different values of $Re$, while the $Re^2/4$ prediction of the model is shown by a 
solid black line. The two calculations agree on nearly two orders of magnitude. 
Positive growth-rate for the large-scale modes implies $\betaMes>1$. 
The critical value of the Reynolds number, for which the instability begins, can 
be obtained graphically at the intersection of the numerically obtained curve with 
the $\betaMes{}=1$ line plotted with a dash-dot green line. The predictions 
of the model $\rec=2$ and the numerically values obtained are in excellent agreement.
%%%%%%%%%%%%%%%%%%%%%%%%%%%%
%%%%%        FIG 11       %%
%%%%%%%%%%%%%%%%%%%%%%%%%%%%
%%%%%%%%%%%%%%%
%%%

\begin{figure}[!ht]
  \centering
  \includegraphics[width=\fwidth,trim= 5 8 5 5, clip=true]{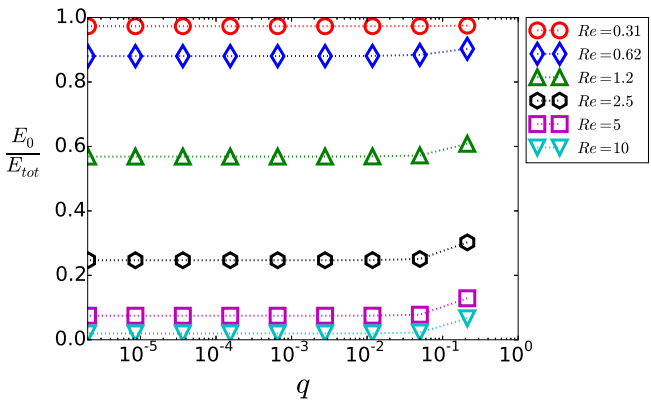}
  \caption{Growth-rate \textit{vs.} Floquet wavenumber, $\sigma(q)$, at different $\rek$ 
  plotted in log-log scale for a Roberts flow.}
  \label{fig:efVqRob}
\end{figure}

\begin{figure}[!ht]
  \centering
  \begin{tikzpicture}
        \node[anchor=south west,inner sep=0] (image) at (0,0) 
        {\includegraphics[width=\fwidth,trim= 20 20 30 20, clip=true]{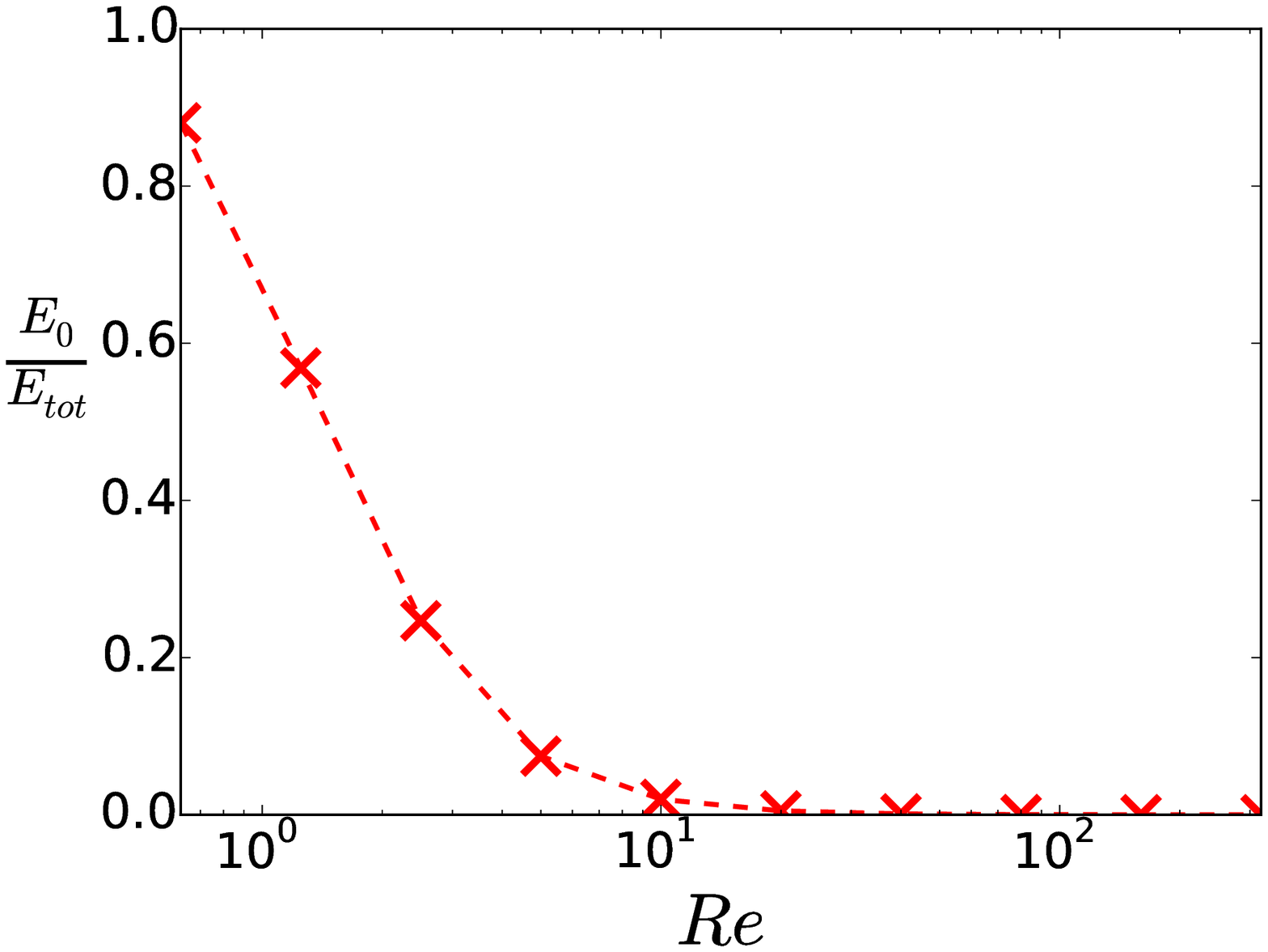}};
        \begin{scope}[x={(image.south east)},y={(image.north west)}]
%            \draw[help lines,xstep=.05,ystep=.05] (0,0) grid (1,1);
%            \foreach \x in {0,1,...,9} { \node [anchor=north] at (\x/10,0) {0.\x}; }
%            \foreach \y in {0,1,...,9} { \node [anchor=east] at (0,\y/10) {0.\y}; }
            \node[anchor=south west,inner sep=0] (image) at (0.35,0.325) 
            {\includegraphics[width=5.4cm,trim= 5 10 35 15, clip=true]{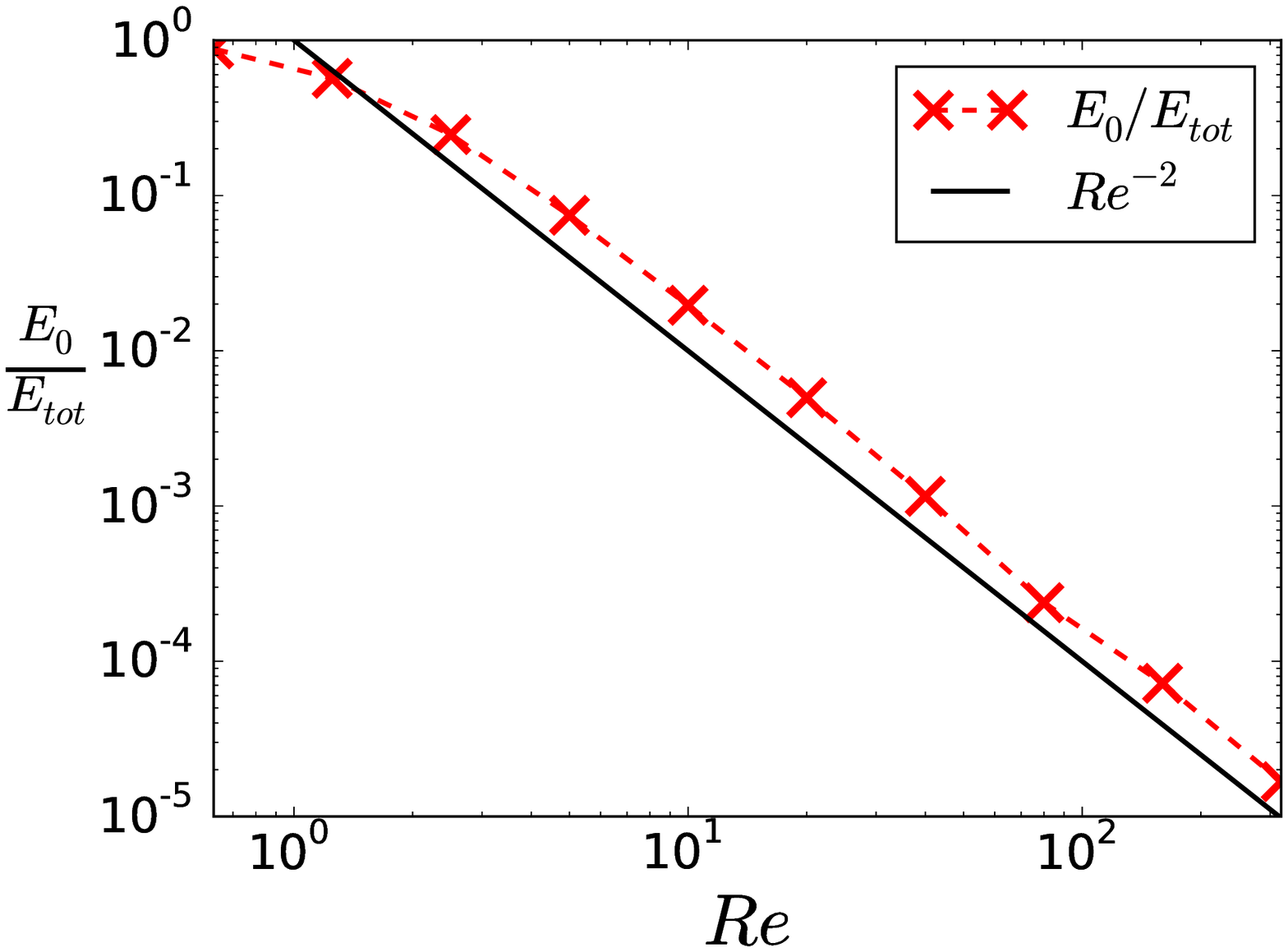}};
        \end{scope}
   \end{tikzpicture}
  \caption{ Fraction of large-scale energy, \EoEt{}, for different 
  \Rn{} for the most unstable mode of the Roberts flow. }
  \label{fig:e0VqVreRob}
\end{figure}

%%%
Similarly to the \AKA{} flow, the fraction of energy concentrated in the 
large scales ($k=1$) becomes independent of $q$ in the small $q$ limit. 
This is demonstrated in fig.~\ref{fig:efVqRob} where the ratio of 
$E_0/E_{tot}$ is plotted as a function of $q$. In fig.~\ref{fig:e0VqVreRob},
we show the asymptotic value of this ratio as a function of the Reynolds number. 
As in the case of the \AKA{} instability, the projection to the large scales 
depends on the Reynolds number, and at large $Re$, it follows the power law 
$\frac{E_0}{E_{tot}} \propto Re^{-2}$.
%%%%%%%%%%%%%%%%%%%%%%%%%%%%
%%%%%        FIG 12       %%
%%%%%%%%%%%%%%%%%%%%%%%%%%%%
%%%%%%%%%%%%%%%
%%%%%%%%%%%%%%%
%%%%%        FIG 13      %%
%%%%%%%%%%%%%%%
%%%%%%%%%%%%%%%
%%%

%%%
%%%%%%%%%%%%%%%%%%%%%%%%%
\subsection{\Dombre{} flow: $\lambda=1$} %%
\begin{figure}[!ht]
  \centering
  \includegraphics[width=\fwidth,trim= 5 5 5 5, clip=true]{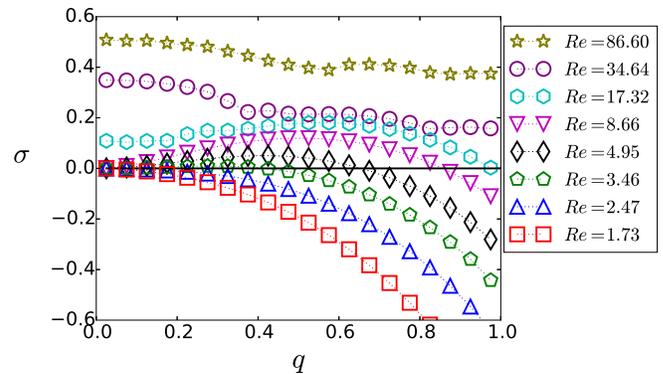}
  \caption{Growth-rate \textit{vs.} Floquet wavenumber $\sigma (q)$, 
  for different $\rek$ for the \ABC{} flow.}
  \label{fig:ABC}
\end{figure}

\label{subsec:Dom}                                        %%
%%%%%%%%%%%%%%%%%%%%%%%%%
For the \dom{} flow, the three-modes model predicts that the \bc{}~coefficient 
is zero. Therefore, the model does not predict a negative eddy-viscosity instability 
with: $\sigma\propto q^2$. Fig.~\ref{fig:ABC} shows the growth-rate as a function of 
the wavenumber $q$ calculated using the Floquet code for different values of the 
Reynolds number. Clearly the small $q$ modes still become unstable but the 
dependence on $Re$ appears different from the previously examined cases.
We thus examine separately the small $Re$ and large $Re$ behaviors.
%%%%%%%%%%%%%%%%%%%%%%%%%%%
%%%%%        FIG 14      %%
%%%%%%%%%%%%%%%%%%%%%%%%%%%
%%%%%%%%%%%%%%%%%%%%%%%%%%%

%%%%%%%%%%%%%%%%%%%%%%%%%%%%%%%%%%%%%%
\subsubsection{Small values of $Re$}%%
\begin{figure}[!ht]
  \centering
  \includegraphics[width=\fwidth,trim= 5 5 5 0, clip=true]{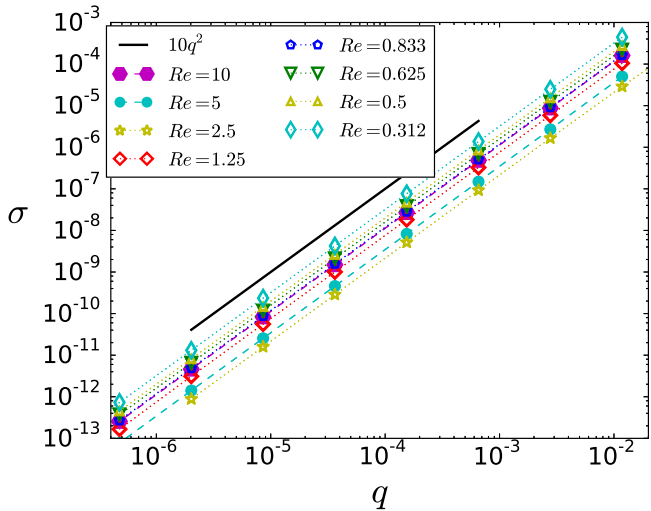}
  \caption{Growth-rate \textit{vs.} Floquet wavenumber, $\sigma(q)$, for different $\rek$ 
  plotted in log-log scale for a \dombre{} flow. The full markers with dashes represent
  the value of positive growth-rates whereas the empty markers with dots represent
  the absolute value of negative growth-rates.}
  \label{fig:sigVqVreDom}
\end{figure}

%%%%%%%%%%%%%%%%%%%%%%%%%%%%%%%%%%%%%% 
First, we examine the instability for small values of $Re\le10$ 
for which the growth-rate $\sigma$ tends to zero as $q\to0$. Fig.~\ref{fig:sigVqVreDom} shows 
the growth-rate of the instability for the \dombre{} flow as a function of the 
wavenumber $q$ in logarithmic scale for different values of $Re$ ranging from $0.312$ to $10$.
In this range, the growth-rate behaves much like the Roberts flow, and is in contradiction with
the three-modes model. The numerically calculated growth-rates show a clear 
negative eddy-viscosity scaling $\sigma \propto q^2$. The growth-rate becomes positive above 
a critical value of $Re$. 
%%%%%%%%%%%%%%%%%%%%%%%%
%%%%%        FIG 15   %%
%%%%%%%%%%%%%%%%%%%%%%%%
%%%%%%%%%%%%%%%%
%%%

\begin{figure}[!htb]
    \centering
    \begin{tikzpicture}
        \node[anchor=south west,inner sep=0] (image) at (0,0) 
        {\includegraphics[width=\fwidth,trim= 5 10 15 5, clip=true]{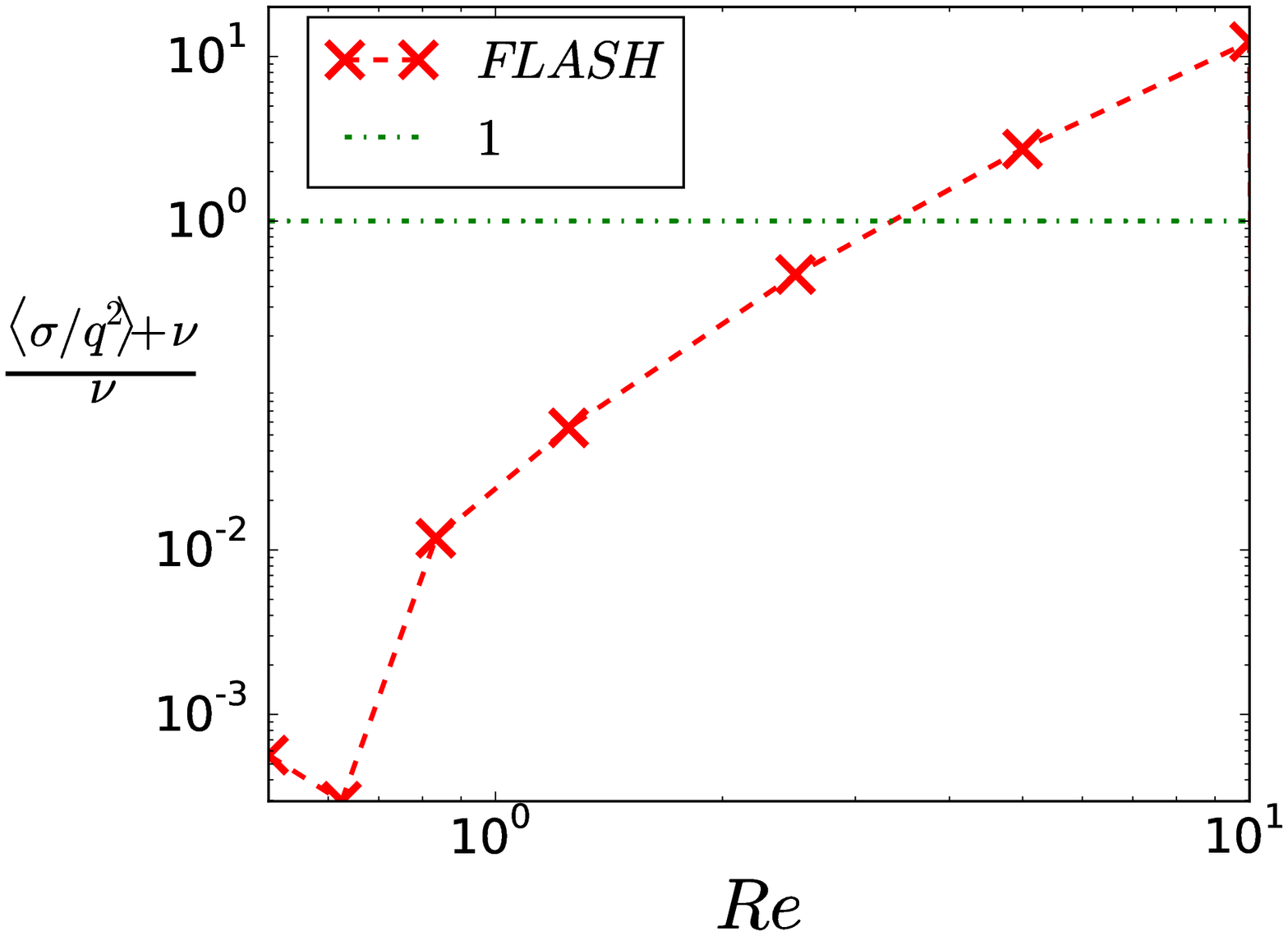}};
        \begin{scope}[x={(image.south east)},y={(image.north west)}]
%            \draw[help lines,xstep=.05,ystep=.05] (0,0) grid (1,1);
%            \foreach \x in {0,1,...,9} { \node [anchor=north] at (\x/10,0) {0.\x}; }
%            \foreach \y in {0,1,...,9} { \node [anchor=east] at (0,\y/10) {0.\y}; }
            \node[anchor=south west,inner sep=0] (image) at (0.52,0.15) 
            {\includegraphics[width=3.9cm,trim= 5 10 15 5, clip=true]{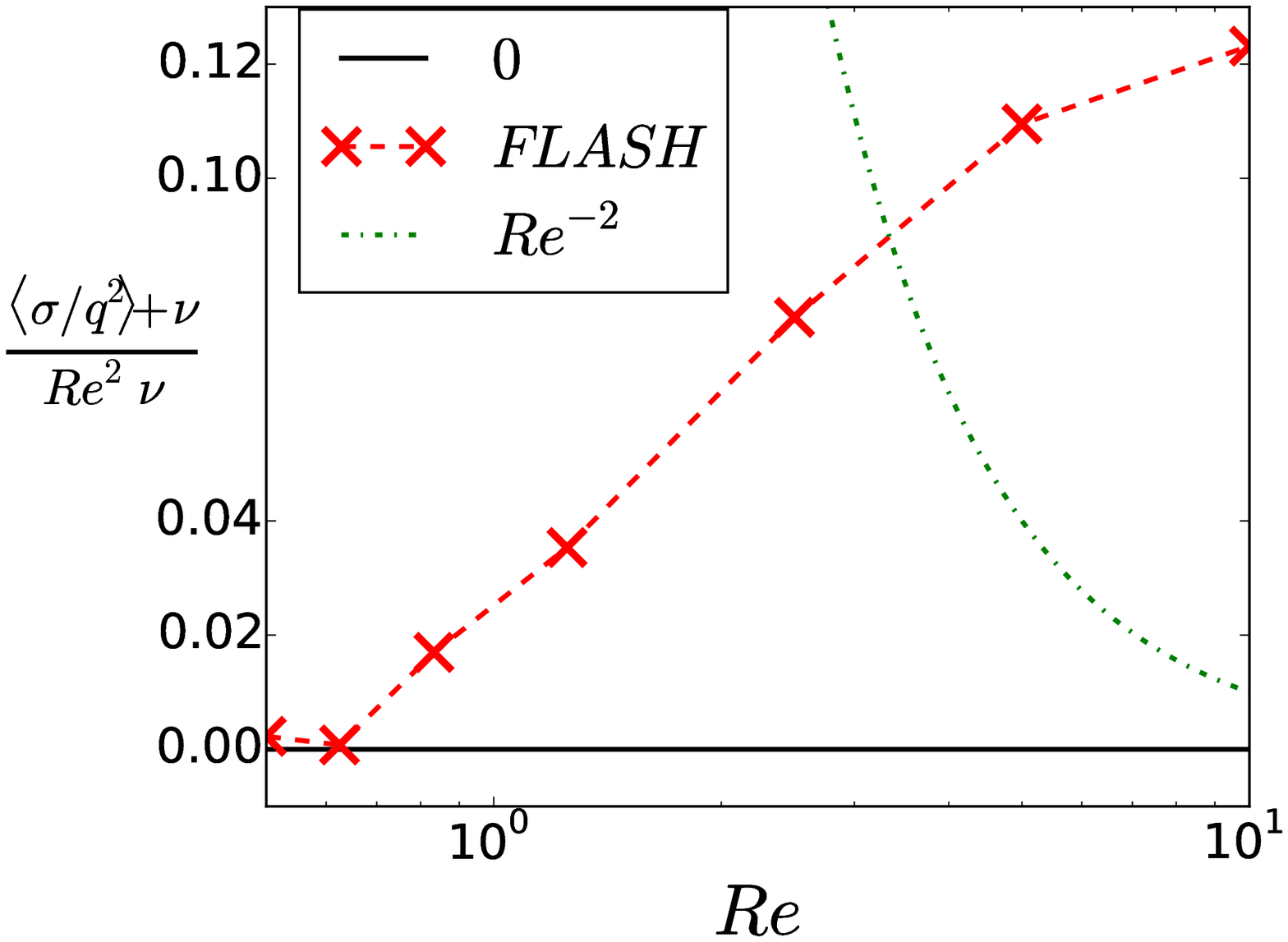}};
        \end{scope}
    \end{tikzpicture}
	\caption{\betaMes{}~coefficient \textit{vs.} \Rn{}, plotted in log-log scale for the \dombre{} flow. 
	The insert shows $b=\betaReMes{}$. }
     \label{fig:betaVreDom}
\end{figure}

%%%
In fig.~\ref{fig:betaVreDom}, the measured value of $\betaMes{}$ is represented 
as a function of the \Rn{}.
In the insert, the plot lin-log of $\betaReMes{}$ provides a measurement of the $b$ coefficient.
This expression becomes larger than one (signifying the instability boundary that is marked 
by a dash-dot line) for $Re \gtrsim 3$. This value $\rec \simeq 3$ is slightly higher than 
the critical Reynolds number of the Roberts flow $\rec=2$. At very small Reynolds number, 
the value of $b=\betaReMes{}$ approaches zero very quickly, which indicates that the model 
prediction is recovered at $Re\to 0$.
%%%%%%%%%%%%%%%%%%%%%%%%%%%%
%%%%%        FIG 16       %%
%%%%%%%%%%%%%%%%%%%%%%%%%%%%
%%%%%%%%%%%%%%%%%
%%%

\begin{figure}[!ht]
  \centering
  \includegraphics[width=\fwidth , trim= 20 25 35 30, clip=true]{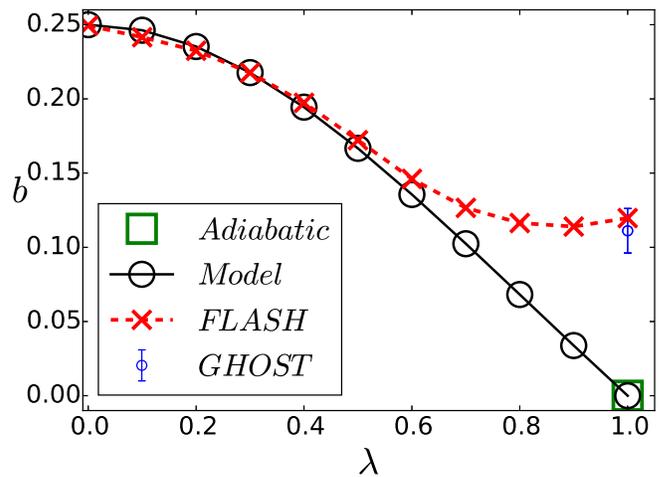}
  \caption{\bc{}~coefficient \textit{vs.} $\lambda$~parameter, $\bc(\lambda)$, at $\rek=10$ 
  for \lamABC{} flows with parameter: $A=1:B=1:C=\lambda\,$.}
  \label{fig:xiVlambda}
\end{figure}

\begin{figure}[!ht]
  \centering
  \includegraphics[width=\fwidth, trim= 20 25 30 30, clip=true]{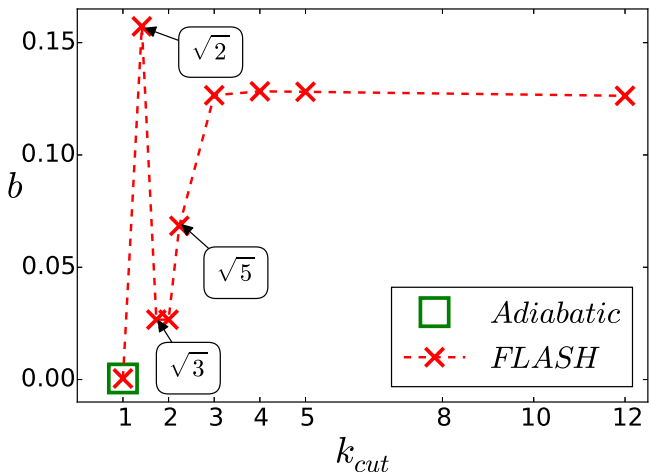}
  \caption{\bc{}~coefficient \textit{vs.} Fourier truncation mode, $\bc(\kcut)$, 
  	at $\rek=10$ of instabilities generated by \lamABC{} flows.}
  \label{fig:xiVkcut}
\end{figure}

%%%
To investigate further the discrepancy of the Floquet results 
with the three-modes model. Fig.~\ref{fig:xiVlambda} shows
the \bc{}~coefficient (measured as $b=\betaReMes{}$) for different $\lambda$~parameter 
from $0$ (Roberts flow) to $1$ (\dombre{} flow). All the DNS are carried out at $\rek=10$. 
%%%%%%%%%%%%%%%
%%%%%        FIG 17      %%
%%%%%%%%%%%%%%%
%%%%%%%%%%%%%%%
The results indicate that the three-modes model and the results from the Floquet code agree 
for $\lambda \lesssim 0.5$ but deviate as $\lambda$ becomes larger.
To identify where this discrepancy between the model and the DNS occurs,
we modified the FLASH code in order to test the assumptions of the model.
This is achieved by enforcing the adiabatic approximation in the Floquet code and
by controlling the number of modes that play a dynamical role.
The later is performed by using a Fourier truncation of the Floquet perturbation at a 
value $k_{cut}$ so that only modes with $k<k_{cut}$ are present. Fig.\ref{fig:xiVkcut} 
shows the dependence of the \bc{}~coefficient on the truncation mode, \kcut{}. 
For $\kcut \geq 3$, the growth-rate reaches the asymptotic value 
that is also observed in the insert of fig.~\ref{fig:betaVreDom} for $Re=10$ obtained 
from the ``untampered" FLASH code. This confirms the assumption that modes in the 
smallest scales have little impact on the evolution of the large-scale perturbation. 
However, the \bc{}~coefficient strongly varies for $\kcut \leq 3$. 
The model predictions are recovered only when $k_{cut}=1$ that amounts to keeping only 
the modes used in the model.
Therefore, the hypothesis of the model to restrict the interaction of the perturbation 
to its first two Fourier modes does not seem to hold for the \dombre{} flow at moderate
\Rn{}, $1\leq \rek \leq 10$. 
The adiabatic hypothesis does not appear to affect the results.
Therefore, the discrepancy between the three-modes model and the numeric results 
is due to the coupling of the truncated velocity $\rmv$ that was neglected in the model.
%%%%%%%%%%%%%%%%%%%%%%%%%%%
%%%%%        FIG 18      %%
%%%%%%%%%%%%%%%%%%%%%%%%%%%
%%%%%%%%%%%%%%%%%%
%%%

%%%
%%%%%%%%%%%%%%%%%%%%%%%%%%%%%%%%%%%%%%
\subsubsection{Large values of $Re$}%%
%%%%%%%%%%%%%%%%%%%%%%%%%%%%%%%%%%%%%%
%%%

\begin{figure}[!ht]
  \centering
  \includegraphics[width=\fwidth , trim= 0 8 12 5 , clip=true]{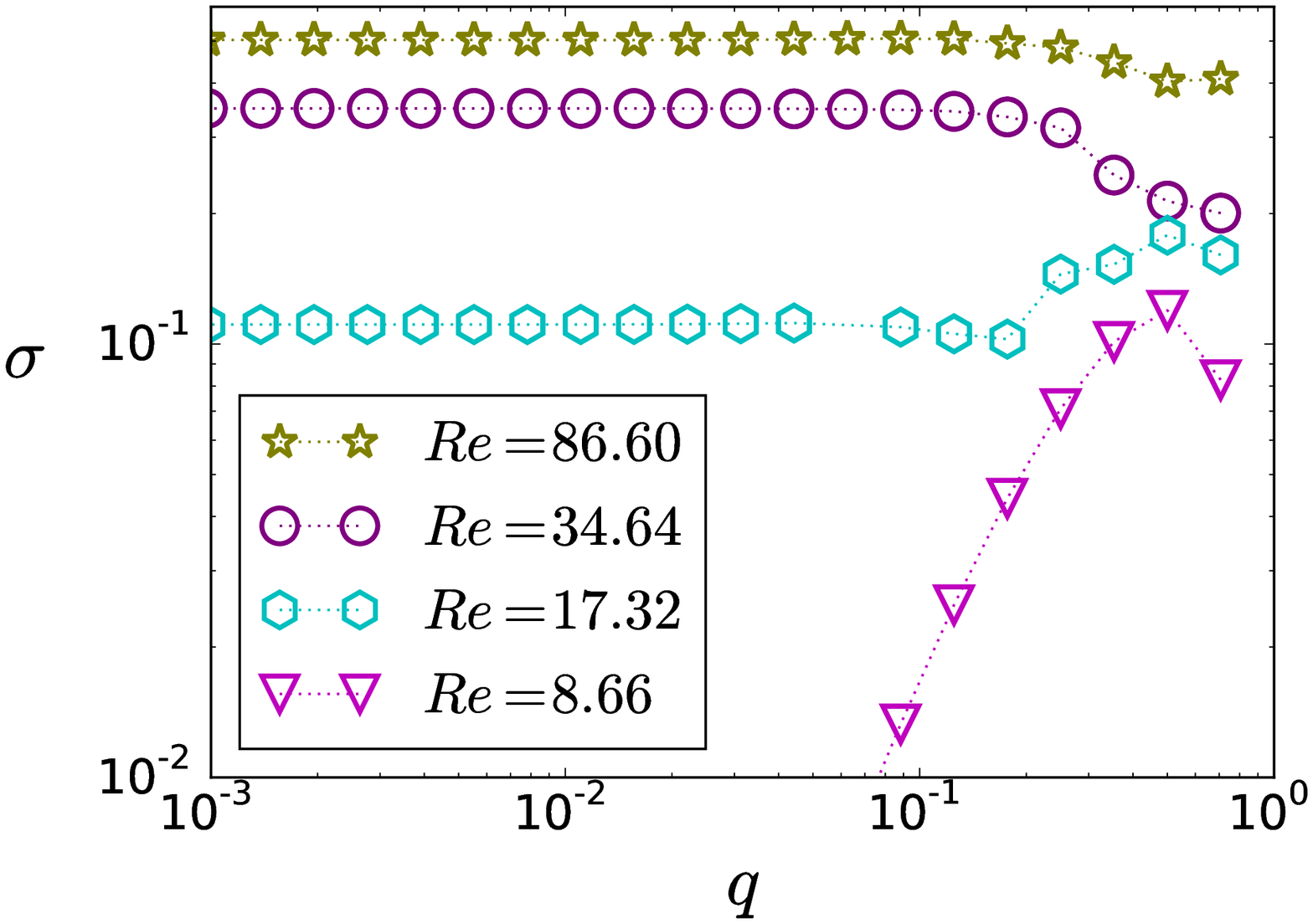}
  \caption{ Growth-rate as a function of $q$ for the \ABC{} 
  flow and for large values of $Re$}
  \label{fig:ABC_highReG}
\end{figure}

%%%
We now turn our focus to large values of the Reynolds number that 
display a finite growth-rate $\sigma$ at $q\to0$, see fig.~\ref{fig:ABC}. 
Fig.~\ref{fig:ABC_highReG} shows the growth-rate $\sigma$ in a lin-log scale 
for four different values of the Reynolds number. Unlike the small values 
of $Re$ examined before here it is clearly demonstrated that 
above a critical value of $Re$ the growth-rate $\sigma$ reaches an asymptotic 
value independent of $q$.
At first, this finite growth-rate seems to violate the momentum conservation.
Indeed, momentum conservation enforces modes with $q=0$, 
corresponding to uniform flows, not to grow.
%%%%%%%%%%%%%%%
%%%%%        FIG 19      %%
%%%%%%%%%%%%%%%
%%%%%%%%%%%%%%%
%%%

\begin{figure}[!ht]
  \centering
  \includegraphics[width=\fwidth , trim= 0 8 5 5 , clip=true]{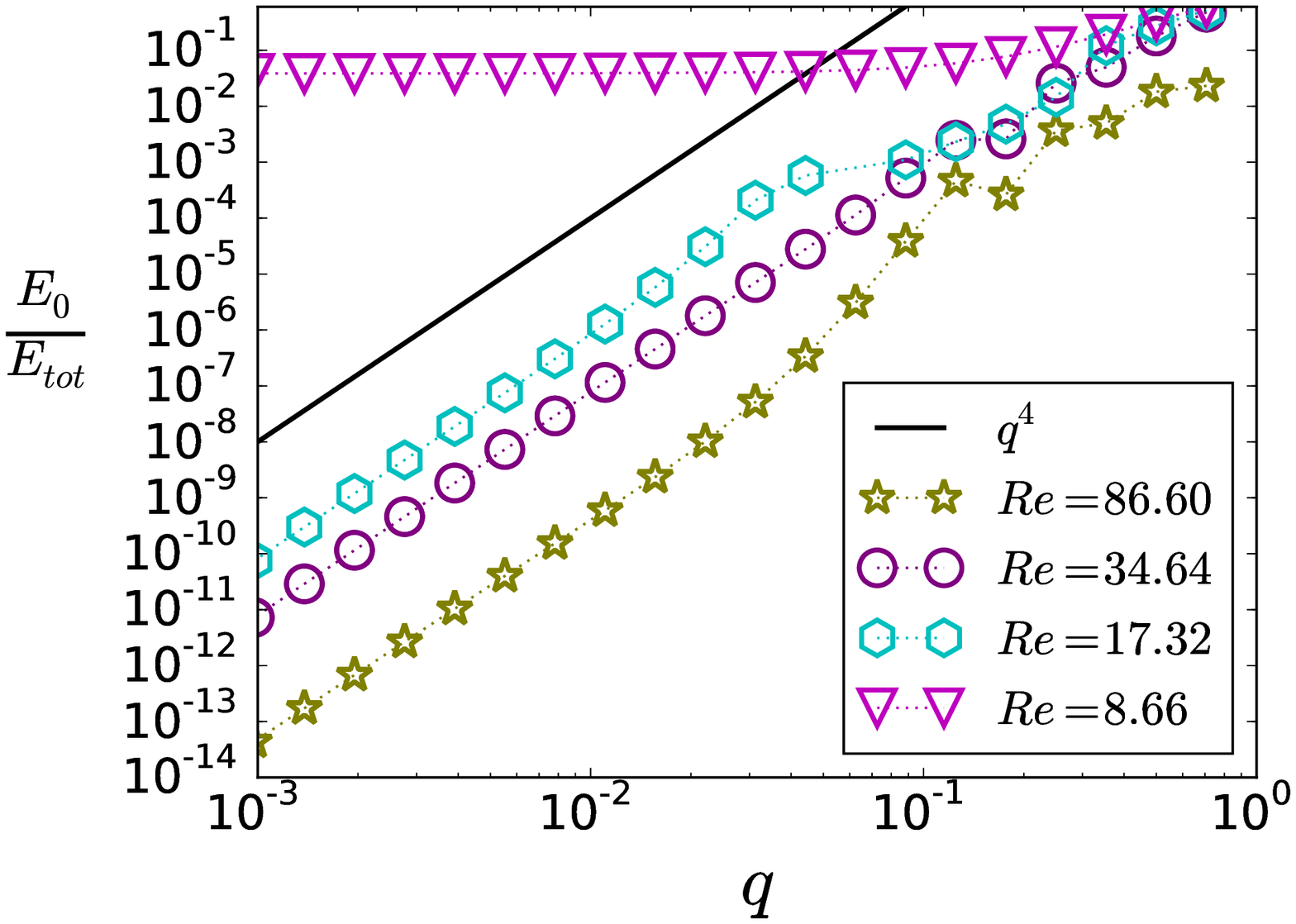}
  \caption{ Growth-rate as a function of $q$ for the \ABC{} 
  flow and for large values of $Re$}
  \label{fig:ABC_E0}
\end{figure}

%%%
The resolution of this conundrum can be obtained by looking at the projection 
of the unstable modes to the large scales. In fig.~\ref{fig:ABC_E0}, we plot the 
ratio $E_0/E_{tot}$ as a function of $q$ for the same values of $Re$ as 
used in fig.~\ref{fig:ABC_highReG}. 
Unlike the small $Re$ cases examined previously, for large $Re$, this energy ratio 
decays to zero at small values of $q$ and appears to follow the power law 
$E_0/E_{tot}\propto q^4$. Therefore, at $q=0$, the energy at 
large scales $E_0$ is zero and the momentum conservation is not violated
in the $q=0$ limit. 
%%%%%%%%%%%%%%%%%%%%%%%%%%%%%%%%%
%%%%%        FIG 20      %%%%%%%%
%%%%%%%%%%%%%%%%%%%%%%%%%%%%%%%%%
%%%%%%%%%%%%%%%%%%%%%%%%%%%%%%%%%
%%%

%%%
%%%%%%%%%%%%%%%%%%%%%%%%%%%%%%%%%%%%%%%%%%%%%%%%%%%%%%%%
\subsubsection{Small and large-scale instabilities}   %%
\label{sec:small}                                                             %%
%%%%%%%%%%%%%%%%%%%%%%%%%%%%%%%%%%%%%%%%%%%%%%%%%%%%%%%%
%%%

%%%
It appears that there are two distinct behaviors: the first one for which
$\lim_{q\to0}\sigma =0$ and $\lim_{q\to0}E_0/E_{tot}>0$ when $Re$ is small and the 
second one for which $\lim_{q\to0}\sigma >0$ and $\lim_{q\to0}E_0/E_{tot}=0$ when 
$Re$ is large. We argue that there is a second critical Reynolds number $\recS$ 
such that flows for which $\rec < Re < \recS$ show the first behavior while 
flows with $\recS < Re$ show the second behavior. This second critical value is 
related to the onset of small-scale instabilities. 
%%%

%%%
To demonstrate this claim we are going to use a simple model. 
We consider the evolution of two modes, one at large scales $v_q$ and one 
at small scales $v_{_Q}$. These modes are coupled together by an external field $U$. 
In the absence of this coupling, the large-scale mode $v_q$ decays while the 
evolution of the small-scale mode $v_{_Q}$ depends on the value of the Reynolds number.
The simplest model satisfying these constraints, dimensionally correct and 
leading to an \AKA{} type $\sigma\propto q$ instability or a negative eddy-viscosity 
instability $\sigma \propto q^2$ is: 
\begin{eqnarray}
	\frac{d}{dt} v_q =& -\nu    q^2 v_q   &+ U q^nQ^{1-n}  v_{_Q} \, ,\\
  \frac{d}{dt} v_Q =&  U Q   v_q        &+ \sigma_{_Q} v_{_Q} \, .
\end{eqnarray}
The index $n$ takes the values $n=1$ if an \AKA{} instability is considered and $n=2$ 
if an instability of negative eddy-viscosity is considered. Note that for $q=0$ 
the growth of $v_q$ is zero, as  required by momentum conservation.
$\sigma_{_Q}=sUQ -\nu Q^2$ gives the small-scale instability growth-rate that is positive 
if $Re=U/(\nu Q)>1/s=\recS$.
%%%

%%%
The simplicity of the model allows for an analytical calculation of the growth-rate 
and the eigenmodes. Despite its simplicity, it can reproduce most of the results 
obtained here in the $q\ll Q$ limit. The general expression for the growth-rate is given by 
$\sigma = \frac{1}{2}\left[(\sigma_{_Q} -\nu q^2 \pm 
\sqrt{ (\sigma_{_Q}+\nu q^2)^2 + 4Q^{2-n}q^nU^2 }\, \right]$ and
eigenmode satisfies $v_q/v_{_Q} =U q^nQ^{1-n}/(\sigma+\nu q^2).$
%%%

%%%
First, we focus on large values of $\nu$ such that $\sigma_{_Q}=-\nu Q^2 <0$.
For $n=1$, the growth-rate $\sigma$ and the energy ratio 
${E_0}/{E_{tot}}= {v_q^2}/{(v_q^2+v_{_Q}^2)}$ are given to the first order in $q$ 
\begin{align}
\sigma \simeq \frac{U^2 q}{\nu Q} \quad \mathrm{and}\quad
\frac{E_0}{E_{tot}} \simeq \frac{1}{1+Re^2}.
\label{eq:mdlA} 
\end{align}
In the same limit for $n=2$ we obtain
\begin{align}
\sigma \simeq \nu (Re^2-1)q^2 \quad \mathrm{and}\quad
\frac{E_0}{E_{tot}} \simeq \frac{1}{1+Re^2}.
\label{eq:mdlB}
\end{align}
The critical Reynolds number for the large-scale instability is given by $\rec=1$.
Both of these results in eqs.~\eqref{eq:mdlA},\eqref{eq:mdlB} are in agreement with the 
results demonstrated in figs.~\ref{fig:sigVqVreAKA}, \ref{fig:specAKAfracq}, 
\ref{fig:specAKAfracRe}, \ref{fig:sigVqVreRob}, \ref{fig:efVqRob}, \ref{fig:e0VqVreRob}.
%%%

%%%
The behavior changes when a small-scale instability exists $\sigma_{_Q}>0$. 
This occurs when $UQ>s\nu Q^2$ at the critical Reynolds number: $\recS=1/s$. 
For large $Re\gg \recS$ we thus expect $\sigma_{_Q} \simeq sUQ>0$.
In this case for $n=1$ to first order in $q$, we have:
\begin{align}
\sigma \simeq \sigma_{_Q} \quad \mathrm{and}\quad
\frac{E_0}{E_{tot}} \simeq \frac{q^2}{s^2Q^2}
\label{eq:mdlC}
\end{align}
while for $n=2$, we obtain: 
\begin{align}
\sigma \simeq \sigma_{_Q} \quad \mathrm{and}\quad
\frac{E_0}{E_{tot}} \simeq \frac{q^4}{s^2Q^4}.
\label{eq:mdlD}
\end{align}
The model is thus in agreement also with the scalings observed in 
figs.~\ref{fig:ABC_highReG},~\ref{fig:ABC_E0}. The transition from one behavior 
to the other occurs at the onset of small-scale instability $\recS$. 
It is thus worth pointing out that the results of the FLASH codes showed 
that the transition from $\lim_{q\to0}\sigma =0$ modes to 
$\lim_{q\to0}\sigma > 0$ occurs at the value of $Re$ for which small-scale instability 
of the \ABC{} flow starts $\recS \simeq13$ \cite{podvigina_non-linear_1994}. 
This further verifies that the transition observed is due to the development of 
small-scale instabilities.
%%%

%%%
We also note here that both the Roberts flow and the $Fr87$ flow given in 
eq.~\eqref{eq:Fr87:flow} are invariant in translations along the $z$-direction. 
This implies that each $q_z$ mode evolves independently with out coupling to 
other $k_z$ modes. The onset of small-scale instabilities $\recS$ for $q=0$ in 
this case then corresponds to the onset of two dimensional instabilities. 
Two dimensional flows however forced at the largest scale of the system 
are known to be stable at all Reynolds numbers \cite{marchioro1986example}. This
result originates from the fact that two dimensional flows conserve both energy 
and enstrophy and small scales cannot be excited without exciting large scales at the 
same time. This is the reason why no $\recS$ were observed in these flows.
%%%

%%%
Finally, this model provides a way to distinguish between the presence or absence of 
the \AKA{} effect for values of $Re$ larger than the critical Reynolds for small-scale 
instabilities $\recS$ by looking at the scaling of the energy in the large scales with 
respect to the scale separation $q/Q$. In the presence of an \AKA{} effect the scaling 
of eq.~\eqref{eq:mdlC} is expected, while, in the absence of an \AKA{} effect, the scaling 
of eq.~\eqref{eq:mdlD} is expected if a negative eddy-viscosity is present.
%%%

%%%
%%%%%%%%%%%%%%%%%%%%%%%
\subsection{Turbulent \dombre{} flows}%%
\label{subsec:turb}                                   %%
%%%%%%%%%%%%%%%%%%%%%%%
%%%

%%%
As discussed in the introduction the driving flow does not need to be laminar to use 
Floquet theory. It is only required to obey the $2\pi \ell$-periodicity. It is worth 
thus considering large-scale instabilities in a turbulent \ABC{} flow that satisfies 
the forcing periodicity. This amounts to the turbulent flow forced by an \ABC{} forcing 
in a periodic cube of the size of the forcing period $2\pi \ell$. Due to the stationarity 
of the laminar \ABC{} flow, it can be excluded as possible candidate for an 
\AKA{} instability. However, this is not true of a turbulent \ABC{} flow since it evolves 
in time. We cannot thus a priori infer that a turbulent \ABC{} flow results in an 
\AKA{} instability or not.
%%%

%%%
To test this possibility, we consider the linear evolution of the large-scale 
perturbations $\vlin$ driven by an \dombre{} flow at $Re=50$, that is beyond the onset of 
the small-scale instability $\recS \simeq 13$. The turbulent \dombre{} flow $\bf U$ is obtained
solving the Navier-Stokes eqs.~\eqref{eq:linNS:K} in the domain $(2\pi \ell)^3$ 
driven by the forcing function ${\bf F}^{ABC} ={\bf U}^{ABC}$.
The code is executed until the flow reaches saturation. The evolution of the large
scale perturbations is then examined solving eq.~\eqref{eq:incomp:Flq} with the FLASH code 
coupled to the Navier-Stokes eqs.~\eqref{eq:linNS:K}. 
%%%

\begin{figure}[!ht]
  \centering
  \includegraphics[width=\fwidth , trim= 0 10 5 5 , clip=true]{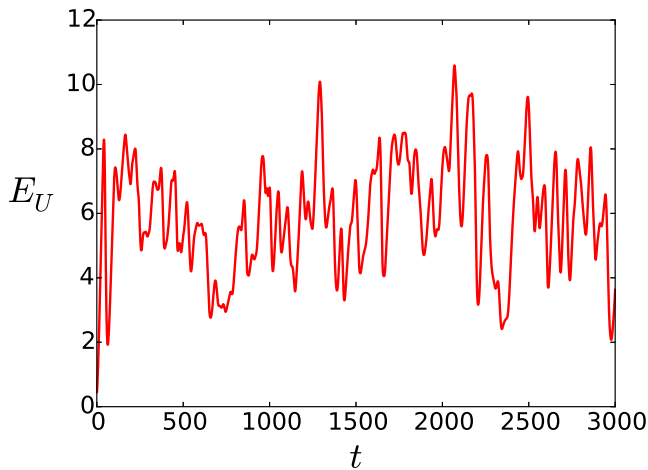}
  \caption{Energy evolution of the turbulent \dombre{} driving flow at $Re=122$.}
  \label{fig:linET_TRB}
\end{figure}

\begin{figure}[!ht]
  \centering
  \begin{tikzpicture}
   \node[anchor=south west,inner sep=0] (image) at (0,0) 
   {\includegraphics[width=\fwidth,trim= 5 5 10 5, clip=true]{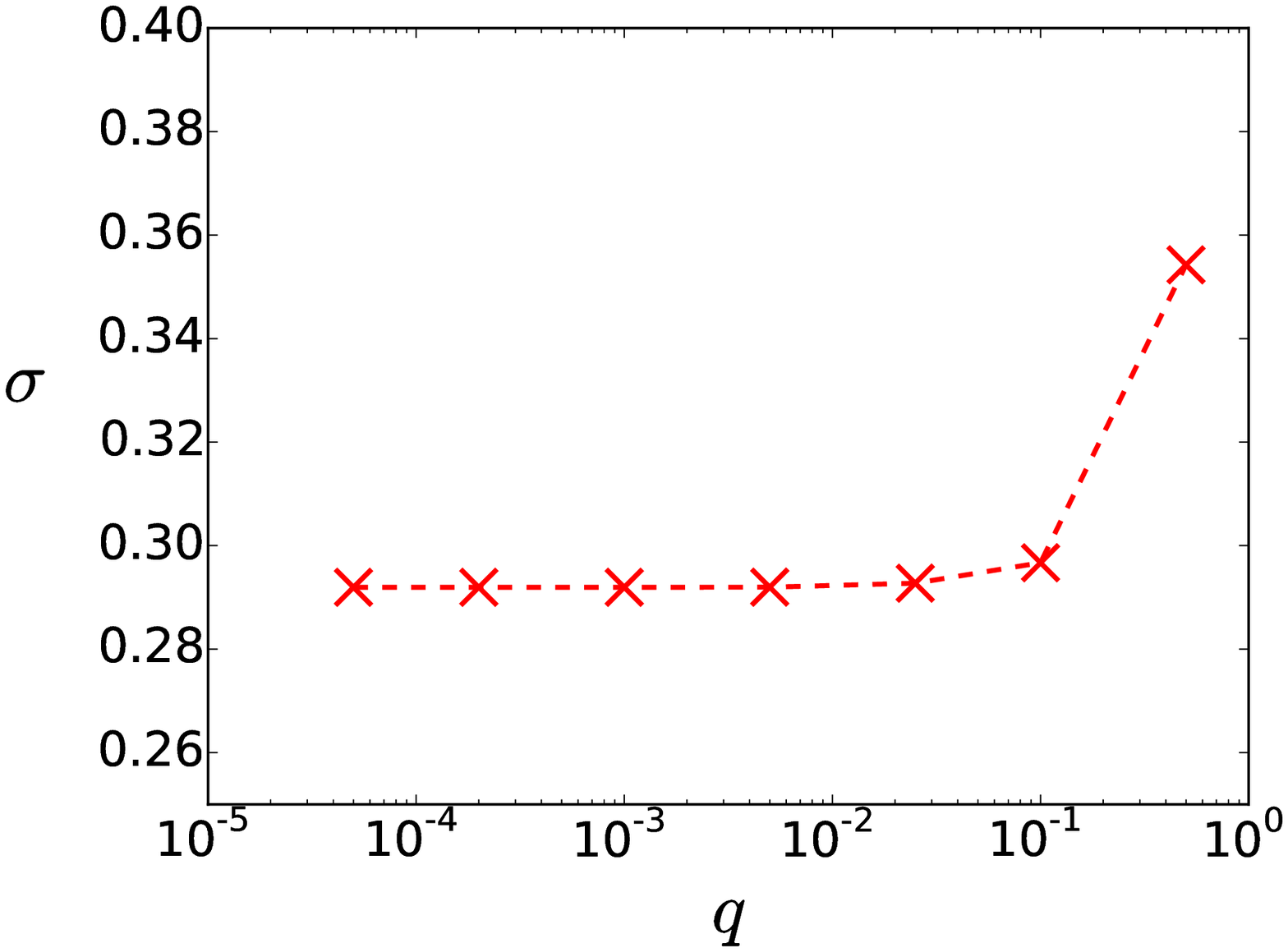}};
     \begin{scope}[x={(image.south east)},y={(image.north west)}]
%            \draw[help lines,xstep=.05,ystep=.05] (0,0) grid (1,1);
%            \foreach \x in {0,1,...,9} { \node [anchor=north] at (\x/10,0) {0.\x}; }
%            \foreach \y in {0,1,...,9} { \node [anchor=east] at (0,\y/10) {0.\y}; }
      \node[anchor=south west,inner sep=0] (image) at (0.2,0.425) 
      {\includegraphics[width=5cm,trim= 5 5 5 5, clip=true]{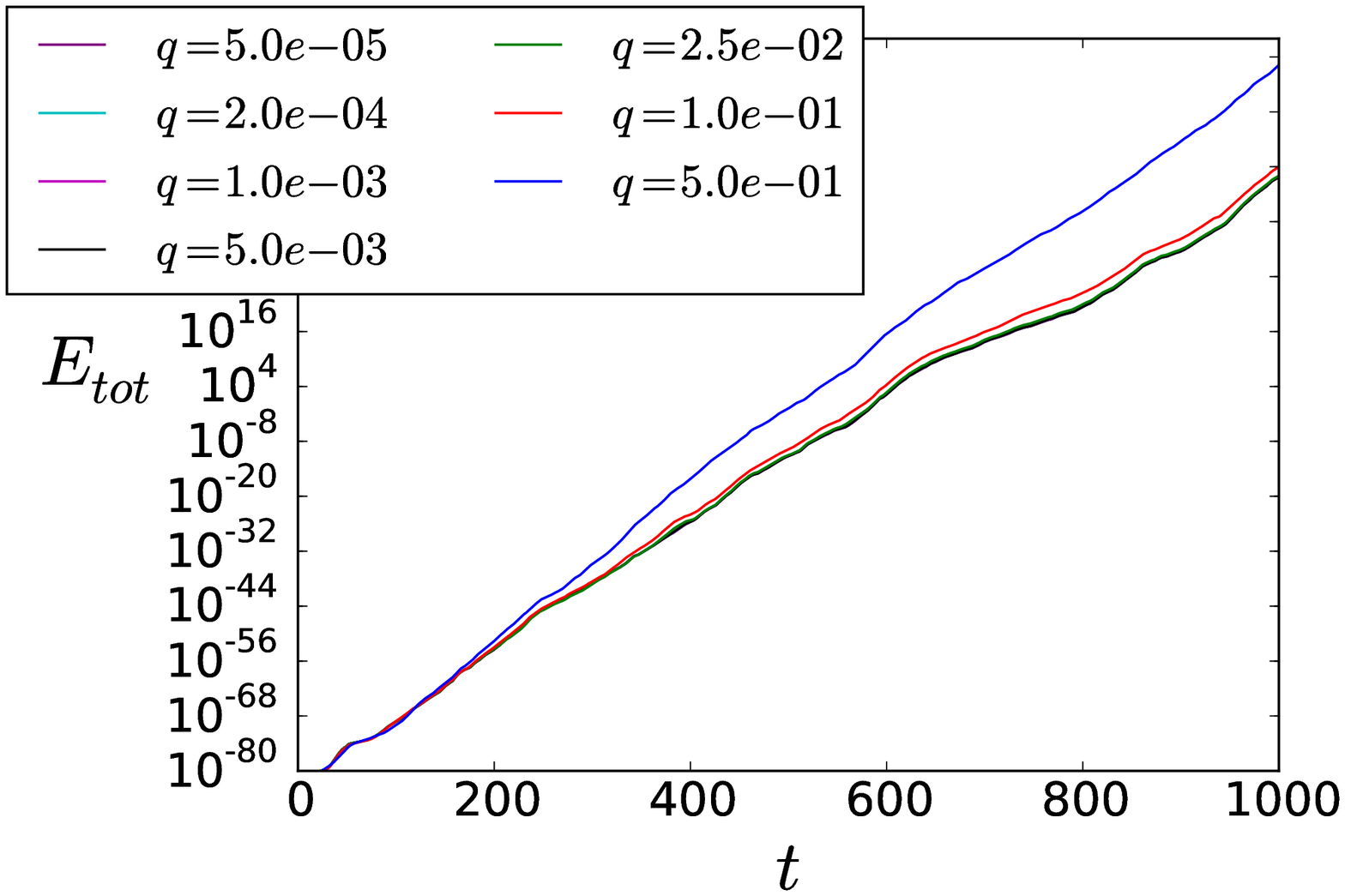}};
    \end{scope}
  \end{tikzpicture}
  \caption{Growth-rate of the turbulent \dombre{} driving flow v, the 
  wavenumber, $\sigma(q)$. The insert shows the exponential growth of 
  the large-scale perturbations for various $q$.}
  \label{fig:TRB_GR}
\end{figure}

\begin{figure}[!ht]
  \centering
  \includegraphics[width=\fwidth]{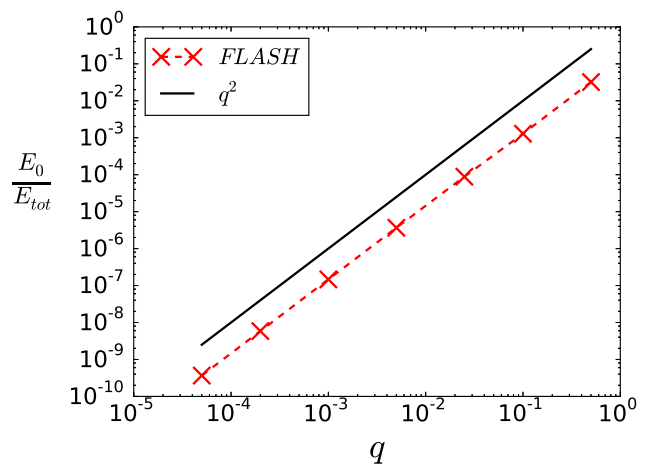}
  \caption{$E_0/E_{tot}$ ratio \textit{vs.} the wavenumber $q$.}
  \label{fig:TRB_E0}
\end{figure}

%%%
The kinetic energy $E_U$ of the turbulent \dombre{} flow $\bf U$ is shown in fig.~\ref{fig:linET_TRB}.
%%%%%%%%%%%%%%%
%%%%%        FIG 21      %%
%%%%%%%%%%%%%%%
%%%%%%%%%%%%%%%%%%%%%%%%%%%%%%%%%
The energy $E_U$ strongly fluctuates around a mean value.
The evolution of the energy $E_{tot}$ of the perturbations $\bf v$ for different values of $q$ 
is shown in the insert of ~\ref{fig:TRB_GR}. $E_{tot}$ shows an exponential increase,
from which the growth-rate can be measured.
The growth-rate $\sigma$ as a function of the wavenumber $q$ is shown in 
fig.~\ref{fig:TRB_GR} while the ratio $E_{0}/E_{tot}$ is shown in fig.~\ref{fig:TRB_E0}.
%%%%%%%%%%%%%%%
%%%%%        FIG 22       %%
%%%%%%%%%%%%%%%
%%%%%%%%%%%%%%%%%%%%%%%%%%%
%%%%%%%%%%%%%%%
%%%%%        FIG 23      %%
%%%%%%%%%%%%%%%
%%%%%%%%%%%%%%%%%%%%%%%%%%%
The growth-rate of the large-scale instabilities appears to reach an finite value in the 
limit $q\to0$ just like laminar $ABC$ flows above the small-scale critical Reynolds $\recS$. 
However, the ratio $E_0/E_{tot}$ does not scale like $q^4$ as laminar \dombre{} flows 
but like $q^2$. As discussed in the previous section, this indicates that the turbulent \dombre{} 
flow is \AKA{}-unstable. This can have possible implications for the saturated stage 
of the instability that we examine next.
%%%

%%%
%%%%%%%%%%%%%%%%%%%%%%%%%%%%%%%%%%%%%
\subsection{Non-linear calculations and bifurcation diagram}        %%
\label{subsec:fullNS}                                                                             %%
%%%%%%%%%%%%%%%%%%%%%%%%%%%%%%%%%%%%%
%%%

%%%
We further pursue our investigation of large-scale instabilities by examining the 
non-linear behavior of the flow close to the instability onset. We restrict ourselves 
to the case of the \dombre{} flow whose non-linear behavior has been extensively studied 
in the absence however of scale separation \cite{dombre_chaotic_1986}. 
The linear stability of the \ABC{} flow in the minimum domain size has been studied in 
\cite{podvigina_non-linear_1994} and more recently in \cite{jones_dynamo_2014}.
These studies have shown that the \ABC{} flow destabilizes at $\recS \simeq 13$. 
%%%

%%%
To investigate the non-linear behavior of the flow in the presence of scale separation, 
we perform a series of DNS of the forced Navier-Stokes equation 
(eq.~\eqref{eq:fullNS}) in triple periodic cubic boxes of size $2\pi L$. The forcing maintaining 
the flow is ${\bf F}^{ABC} = \frac{\sqrt{2}}{\sqrt{3}} \nu |{\bf K}|^2 {\bf U}^{ABC}$ so
that the laminar solution of the flow is the \ABC{} flow \cite{dombre_chaotic_1986}
normalized to have unit energy.
Four different boxes sizes are considered: $KL=1,5,10$ and $20$. For each box size 
and for each value of $Re$, the flow is initialized with random initial conditions and 
evolves until a steady state is reached.
%%%

\begin{figure}[!htb]
    \centering
    \begin{tikzpicture}
        \node[anchor=south west,inner sep=0] (image) at (0,0) 
        {\includegraphics[width=\fwidth,trim= 35 25 25 25, clip=true]{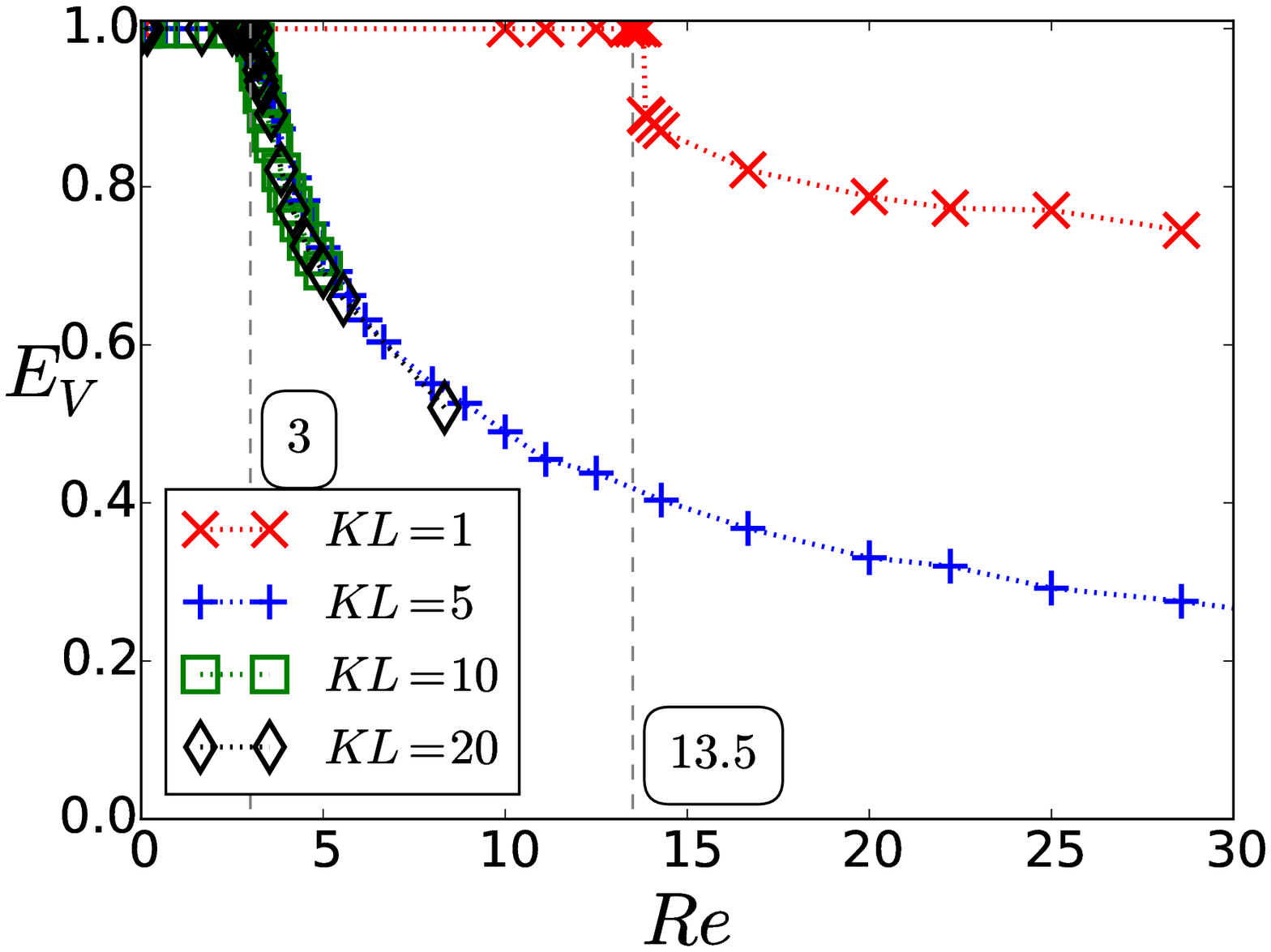}};
        \begin{scope}[x={(image.south east)},y={(image.north west)}]
%            \draw[help lines,xstep=.05,ystep=.05] (0,0) grid (1,1);
%            \foreach \x in {0,1,...,9} { \node [anchor=north] at (\x/10,0) {0.\x}; }
%            \foreach \y in {0,1,...,9} { \node [anchor=east] at (0,\y/10) {0.\y}; }
            \node[anchor=south west,inner sep=0] (image) at (0.52,0.27)
            {\includegraphics[width=3.85cm,trim= 25 25 25 25, clip=true]{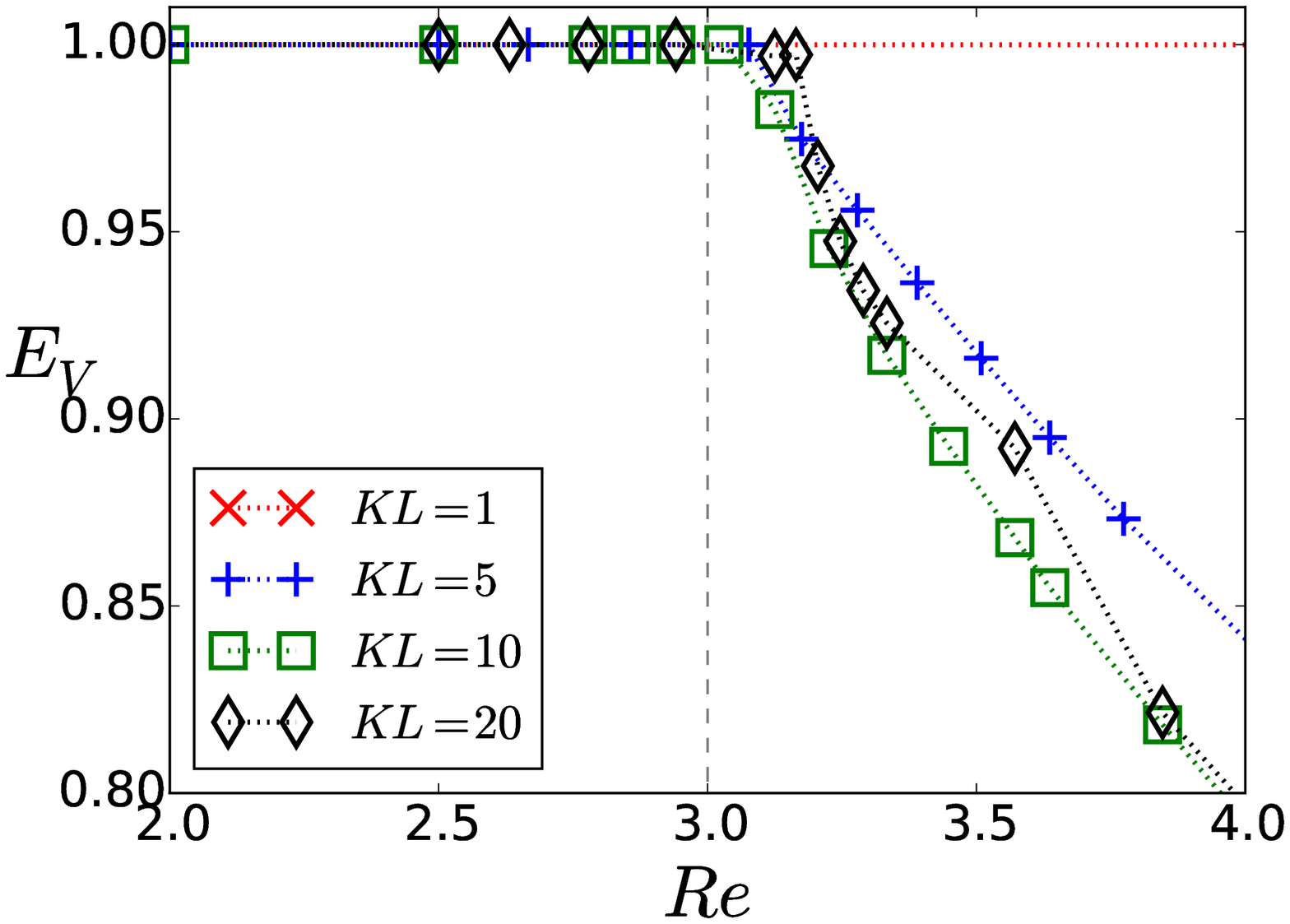}};
        \end{scope}
    \end{tikzpicture}
	\caption{Bifurcation: total energy \textit{vs.} \Rn{}, 
	$E_{tot}(Re)$, for different scale separation 
	$K\in \lbrace 1 ; 5; 10 ; 20 \rbrace $. In insert, zoom of the graph
	for $\rek \in [ 2;5 ]$.}
     \label{fig:NrgVreK}
\end{figure}

%%%
Fig.~\ref{fig:NrgVreK} shows the saturation level of the total energy $E_V$ at steady state 
as a function of $Re$ for the four different values of $KL$.
At low \Rn{}, the laminar solution ${\bf V=U}^{ABC}$ is the only attractor and so the 
energy is $E_V=1$. 
At the onset of the instability the total energy decreases.
A striking difference appears between the $KL=1$ case and other three cases.
For the $KL=1$ case the first instability appears at $\recS \simeq 13$ in agreement 
with the previous work \cite{podvigina_non-linear_1994,jones_dynamo_2014}.
By definition, only small-scale instabilities are present in the $KL=1$ case
({\it i.e.} instabilities that do not break the forcing periodicity).
For the other three cases, which allow the presence of modes of larger scale than the
forcing scale, the flow becomes unstable at a much smaller value: $\rec \simeq 3$.
This value of $\rec$ is in agreement with the results obtained in section \ref{subsec:Dom}
for large-scale instability by a negative eddy-viscosity mechanism.
%%%%%%%%%%%%%%%%%%%%%%%%%%%
%%%%%%%%%%%%%%%
%%%%%        FIG 24      %%
%%%%%%%%%%%%%%%
%%%%%%%%%%%%%%%%%%%%%%%%%%%
The energy curves for the forcing modes $KL \geq 5$ all collapse on the same curve. 
This indicates that not only the growth-rate but also the saturation mechanism for these three 
simulations are similar.
%%%

\begin{figure}[!htb]
    \centering
    \begin{tikzpicture}
    \node[anchor=south west,inner sep=0] (image) at (0,0) 
    {\includegraphics[width=\fwidth]{fig25a}};
        \begin{scope}[x={(image.south east)},y={(image.north west)}]
%            \draw[help lines,xstep=.05,ystep=.05] (0,0) grid (1,1);
%            \foreach \x in {0,1,...,9} { \node [anchor=north] at (\x/10,0) {0.\x}; }
%            \foreach \y in {0,1,...,9} { \node [anchor=east] at (0,\y/10) {0.\y}; }
        \node[anchor=south west,inner sep=0] (image) at (0.0,0.5) 
        {\includegraphics[width=\hwidth,trim= 0 5 5 5, clip=true]{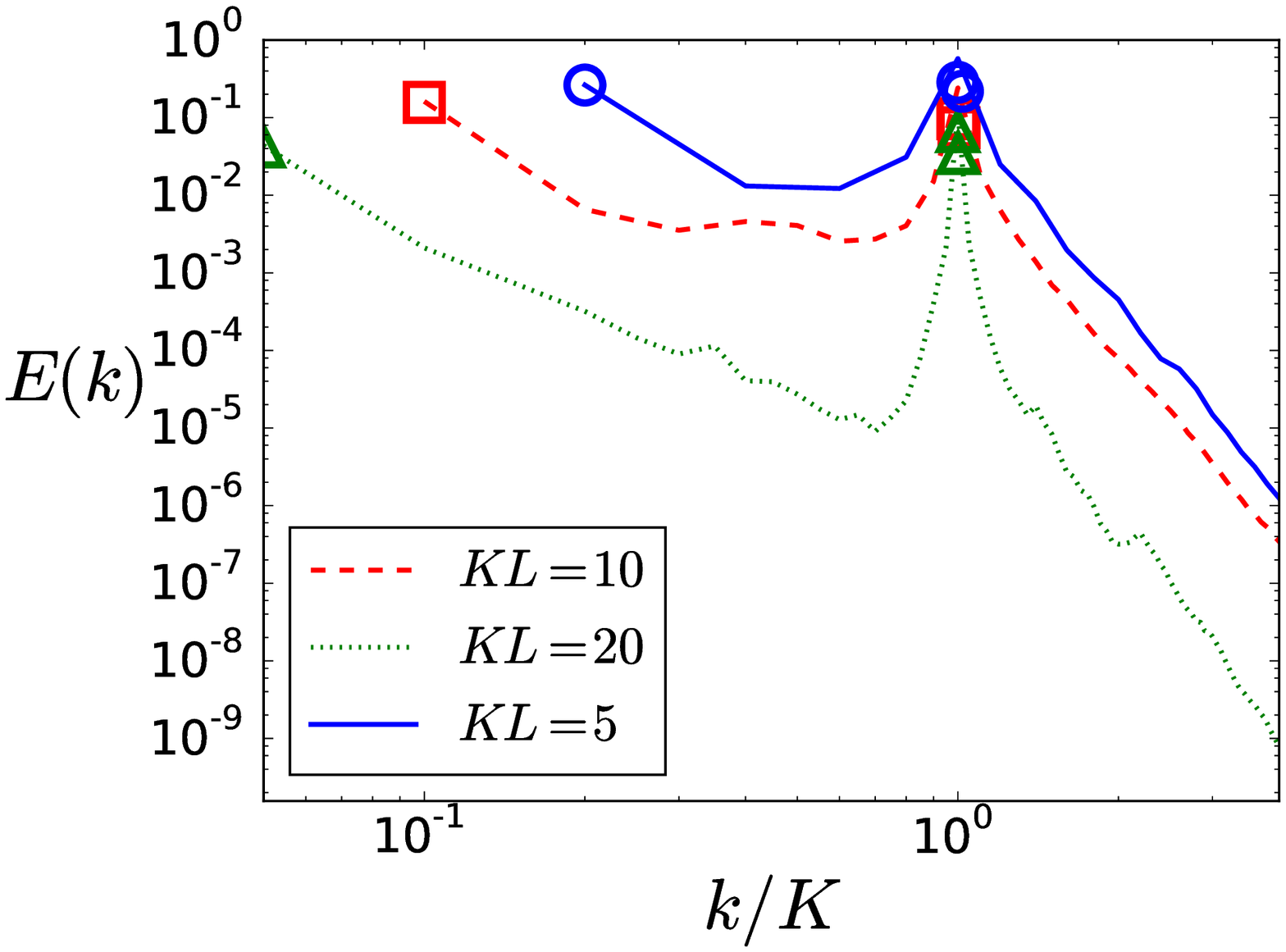}};
        \node[anchor=south west,inner sep=0] (image) at (0.5,0.5) 
        {\includegraphics[width=\hwidth,trim= 0 5 5 5, clip=true]{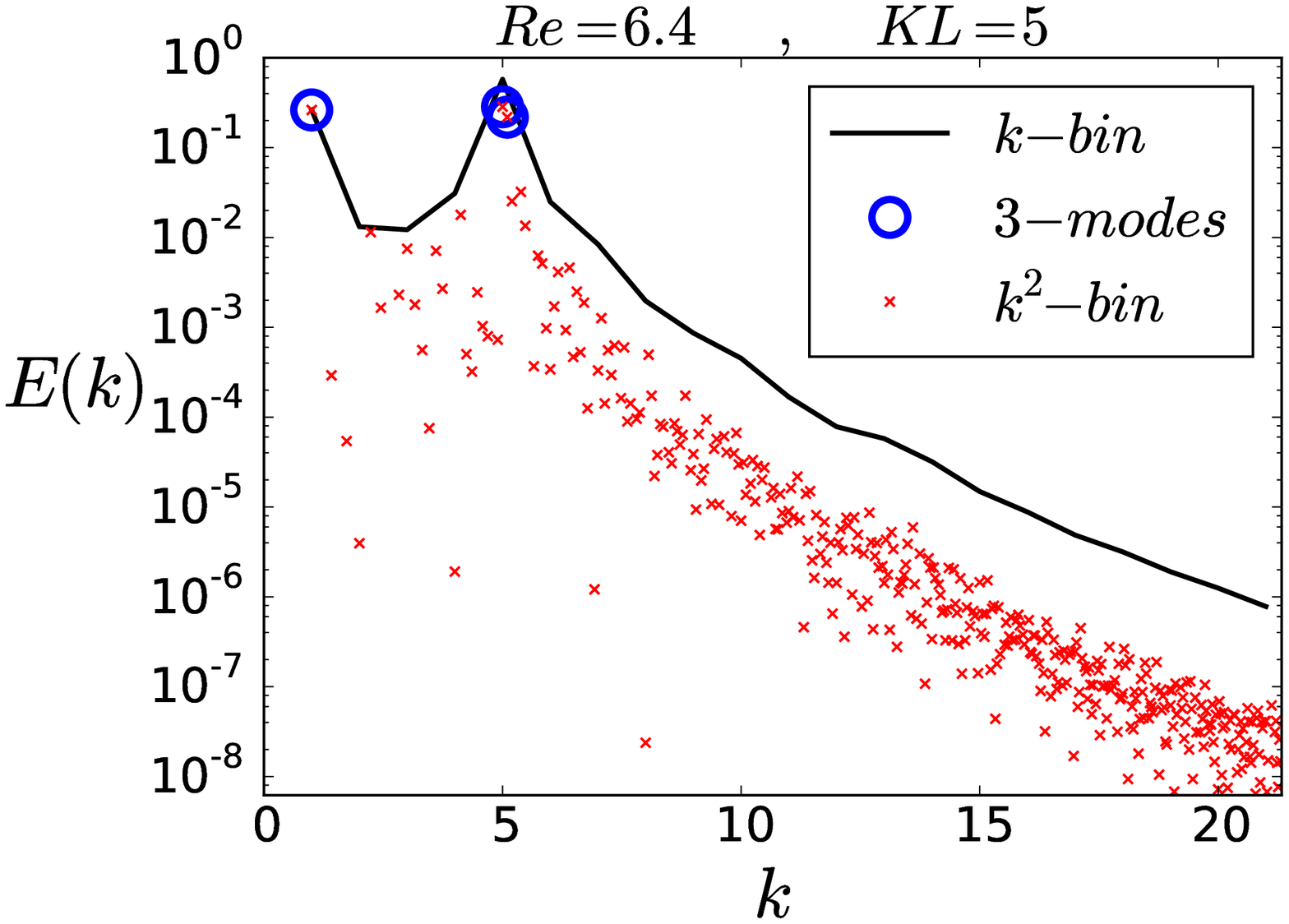}};
        \node[anchor=south west,inner sep=0] (image) at (0.0,0.0) 
        {\includegraphics[width=\hwidth,trim= 0 5 5 5, clip=true]{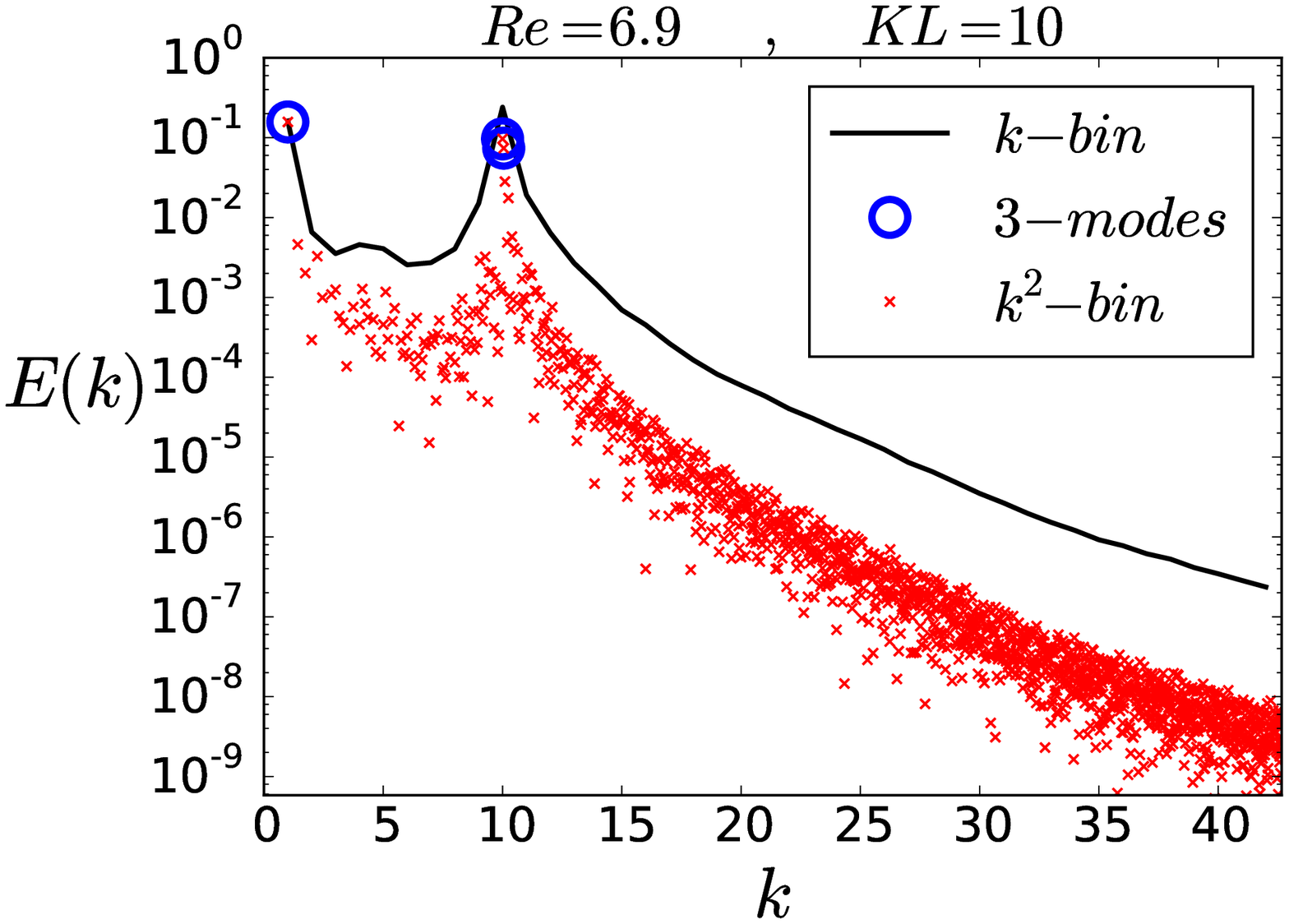}};
        \node[anchor=south west,inner sep=0] (image) at (0.5,0.0) 
        {\includegraphics[width=\hwidth,trim= 0 5 5 5, clip=true]{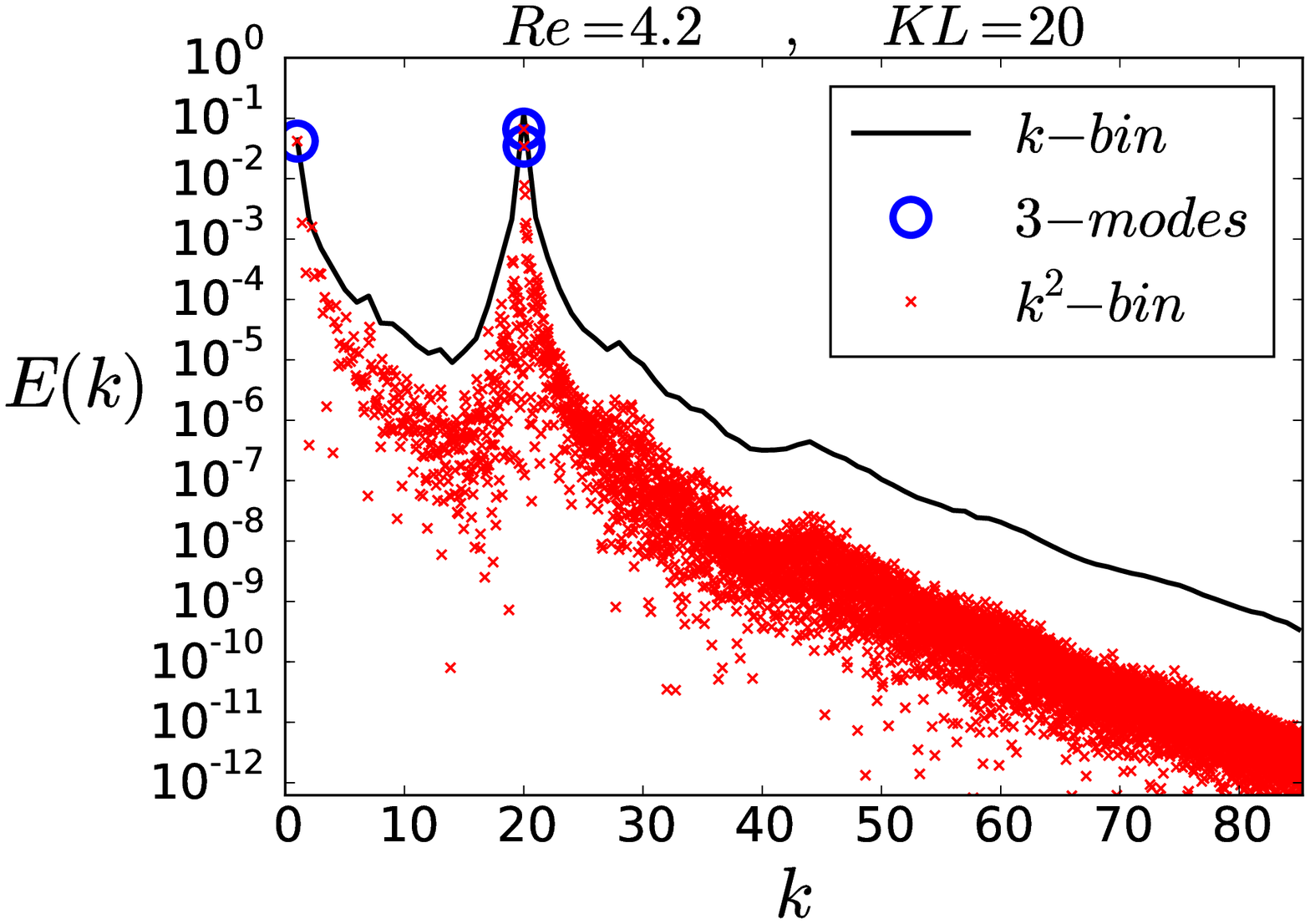}};
    \end{scope}
    \end{tikzpicture}
	\caption{Energy spectra, $E(k)$, for different scale separation 
	$K\in \lbrace 1 ; 5; 10 ; 20 \rbrace $.}
     \label{fig:Ek}
\end{figure}

%%%
%%%%%%%%%%%%%%%%%%%%%%%%%%%
Further insight on the saturation mechanism can be obtained by looking at the energy spectra.
Fig.~\ref{fig:Ek} shows the energy spectrum of the velocity field at the steady state of the simulations.
Two types of spectra are plotted. In fig.~\ref{fig:Ek}, spectra plotted using lines and 
denoted as $k$-bin display energy spectrum collected in bins where modes $\bf k$ satisfy 
$ n_1-1/2 < {|\bf k}|L \leq n_1+1/2$, with $n_1$ a positive integer.
$E(k)$ then represents the energy in the bin $n_1=k$.
In fig.~\ref{fig:Ek}, spectra plotted using red dots and denoted by $k^2\!$-bin display the
energy spectrum collected in bins where modes $\bf k$ satisfy $ |{\bf k}|^2L^2=n_2$, 
with $n_2$ a positive integer. Since ${\bf k}L$ is a vector with integer components $m_x$,
$m_y$ and $m_z$, its norm $k^2L^2=m_x^2+m_y^2+m_z^2$ is also
a positive integer. $E(k)$ then represents the energy in the bin $n_2=k^2L^2$.
This type of spectrum provides more precise information about the energy distribution among modes. 
In our case, they help separate $K$ modes from $K\pm1/L$ modes and highlight the three-modes
interaction. The $k=K\pm1/L$ modes as well as the largest scale mode $kL=1$ that were used 
in the three-modes model are shown by blue circles in the spectra.
The drawback of $k^2\!$-bin spectra is their memory consumption. They have a
number of bins equal to the square of the number of bins of standard $k$-bin spectra.
However, since spectra are not outputted at every time-step, this inconvenience is
limited.
%%%

%%%%%%%%%%%%%%%%%%%%%%%%%%%%%%
%%%%%%%%%%%%%%%
%%%%%        FIG 25      %%
%%%%%%%%%%%%%%%
%%%%%%%%%%%%%%%%%%%%%%%%%%%
%%%

%%%
The plots of the spectra show that the most energetic modes are 
the modes close to the forcing scale and the largest scale mode $kL=1$. This 
is true even for the largest scale separation examined $KL=20$. We note that the largest
scale mode is not the most unstable one as seen in all the cases examined 
(see figs.~\ref{fig:sigVqVreAKA_lin},\ref{fig:linRob},\ref{fig:ABC}).
Despite this fact, it appears that the $kL=1$ is the dominant mode that 
controls saturation. The exact saturation mechanism however is beyond the scope of this work. 
%%%

%%%
\section{Conclusion}
\label{sec:ccl}
%%%

%%%
In this work, we examined in detail the large-scale hydrodynamic instabilities 
of a variety of flows. Using the Floquet framework as well as simplified models, 
we were able to investigate the stability of periodic flows to large-scale perturbations 
for a wide parameter range. Our work verifies the asymptotic results derived in 
the past but also covers cases that go beyond their validity including turbulent flows. 
%%%

%%%
For the $Fr87$ flow (see eq.~\eqref{eq:Fr87:flow}) at small values of $Re$, the instability growth 
rate scales like: $\sigma \propto q\; Re\,$, with most of the energy in the large scales
$1-E_0/E_{tot} \propto Re^2$. It is present for any arbitrarily small value of the 
Reynolds number provided that scale separation is large enough. When $Re$ becomes of order 
one this behavior changes. The growth-rate saturates in $Re$ and most of the energy of the 
most unstable mode is concentrated in the small scales. 
%%%

%%%
Flows in the absence of an \AKA{} effect, like the \ABC{} and Roberts flow, show a 
negative eddy-viscosity scaling. The instability appears only above a critical value 
of the Reynolds number $\rec$ that was found to be $\rec \simeq 2$ for the Roberts flow
and $\rec \simeq 3$ for the \dombre{} flow. The growth-rate follows the scaling 
$\sigma \propto \nu(b Re^2-1) q^2$. The value of $b$ can be calculated based on a three 
mode model for the Roberts flow and was found to be $b=1/4$. The three-modes model however 
failed to predict the \bc{}~coefficient of the \dombre{} flow because more modes were contributing 
to the instability.
For the \dombre{}, the negative eddy-viscosity instability was shown to stop at a second critical 
Reynolds number $\recS \simeq 13$, where the flow becomes unstable to small-scale perturbations. 
For values of $Re$ larger than $\recS$ the growth-rate remains finite and independent of $q$ even at 
the $q\to 0 $ limit. On the contrary, the fraction of energy at the largest scale becomes
dependent on $q$ decreasing as $E_0/E_{tot} \propto q^4$ in the $q\to0$ limit. These 
behavior is well described by a two-modes model that is explained in sec.~\ref{sec:small}. 
This model also predicts that in the case of an \AKA{} 
instability the ratio $E_0/E_{tot}$ scales like $E_0/E_{tot} \propto q^2$ for $Re>\recS$.
This scaling was indeed found by examining the large-scale instability of a turbulent ABC flow,
indicating that a turbulent \dombre{} is \AKA{}-unstable.
%%%

%%%
Our study was carried out further to the non-linear regime where it was shown that 
in the presence of scale separation, the forcing scale and the largest scales of the system are 
the most dominant energetically. The persistence of this behavior at larger values of $Re$ 
remains to be examined.
%%%

%%%
\acknowledgments 
%%%

%%%
This work was granted access to the HPC resources of MesoPSL financed by the Region 
Ile de France and the project Equip@Meso (reference ANR-10-EQPX-29-01) of 
the programme Investissements d'Avenir supervised by the Agence Nationale pour 
la Recherche and the HPC resources of GENCI-TGCC-CURIE \& GENCI-CINES-JADE 
(Project No. x20162a7620) where the present numerical simulations have been performed.
%%%

%%%
\section{Appendix: FLASH}
\label{sec:apx:FLASH}
A pseudo-spectral method is adopted to compute numerically eq.~\eqref{eq:ns:Flq} 
and \eqref{eq:incomp:Flq}. The linear term are computed in Fourier space. All the 
terms involving the driving flow are computed in physical space made incompressible 
by solving in periodic space the Poisson problem, using:
\begin{align}
	\bm{\Psi}^{(2)} = -\Laplace^{-1} (\roT)^2 \bm{\Psi}^{(1)} \,.
\end{align}
The main steps of the algorithm are written below.
In this algorithm, \Four{} and $\Four^{-1}$ denote direct and inverse
fast Fourier transforms. $\vect{AUX}^{(1)}$ and $\vect{AUX}^{(2)}$ are two 
auxiliary vector fields. $\vect{AUX}^{(1)}$ is real and $\vect{AUX}^{(2)}$ is complex.
\begin{algorithm}[!htb]
\caption*{Floquet Linear Analysis of Spectral Hydrodynamic (FLASH)}
\begin{algorithmic}[1]
\REQUIRE $\nu$, $T$, $dt$, $\qvec$, $\vfloq{}^{(0)}$, $\Ulam$ 
\STATE $\Wlam= \roT \Ulam$
\STATE $n=0$
\STATE $\vect{V}^{(n)}=\Four(\vlin{}^{(n)})$
\WHILE{$t < T$}
	\STATE $\vect{AUX}^{(1)}= \Ulam \times \Four^{-1}
	(\imath ( \kvec + \qvec) \times \vect{V}^{(n)}) 
	- \Wlam \times \Four^{-1} (\vect{V}^{(n)}) $
%	C = C-D
	\STATE $\vect{AUX}^{(2)}=- \vert\vert \kvec + \qvec\vert\vert^{-2} 
	( \kvec + \qvec) \times ( \kvec + \qvec) \times \mathcal{F} [\vect{AUX}^{(1)}]$ 
%	B = invlap rot rot C	
	\STATE $\vect{V}^{(n+1)} = \vect{V}^{(n)} + 
	dt ( \vect{AUX}^{(2)} - \nu \vert\vert \kvec + \qvec\vert\vert^{2} \vect{V}^{(n)} )$
%	C = A + dt ( B + nu J )
	\STATE $n=n+1$ , $t=t+dt$ 
\ENDWHILE
\end{algorithmic}
\end{algorithm}
%%%

%%%
To carry out the computations with greater precision, a fourth order Runge-Kutta
method is used instead of the simple Euler method at line 7 of the algorithm. 
The Fourier parallel expansions are also truncated at $1/3$ to avoid aliasing error. 
The code is parallelised with MPI and uses many routine from the GHOST code 
\cite{mininni_nonlocal_2008}.
Most of the DNS are done at a $32^3$ and $64^3$ resolution. Convergence tests show
that this resolution is sufficient for the range of \Rn{} studied.
%%%

%%%
\bibliography{vUV}
%%%

%%%
\end{document}